\documentclass[prb,preprintnumbers,amsfonts,amssymb,amsmath,floats,twocolumn,aps]{revtex4}

\usepackage[pdftex]{graphicx}
\usepackage{dcolumn}
\usepackage{bm}
\usepackage{color}

\usepackage{amsmath}
\usepackage{amssymb}
\usepackage{todonotes}
\usepackage{graphicx}
\usepackage{physics}     
\usepackage{tcolorbox}      

\usepackage[pdftex,colorlinks=true,linkcolor=blue,citecolor=blue,filecolor=blue]{hyperref}

\newcommand{\ff}[1]{{\boldsymbol #1}}
\newcommand{\ca}[1]{{\cal #1}}
\newcommand{\bi}{\begin{itemize}}
\newcommand{\ei}{\end{itemize}}
\newcommand{\be}{\begin{equation}}
\newcommand{\ee}{\end{equation}}
\newcommand{\ba}{\begin{eqnarray}}
\newcommand{\ea}{\end{eqnarray}}

\newcommand{\jd}{J^{\rm (dim)}_{\rm c}}
\newcommand{\jm}{J^{\rm mag}_{\rm c}}
\newcommand{\jdv}{0.89t_{1}}
\newcommand{\jdva}{0.62t}
\newcommand{\jmv}{0.84t_{1}}

\begin{document} 
  
\title{Phase diagram of the Kondo model on the zigzag ladder}
\author{Matthias Peschke, Lena-Marie Woelk and Michael Potthoff}
\affiliation{Fachbereich Physik, Universit\"at Hamburg, Jungiusstra\ss{}e 9, 20355 Hamburg, Germany}

\begin{abstract}
The effect of next-nearest-neighbor hopping $t_{2}$ on the ground-state phase diagram of the one-dimensional Kondo lattice is studied with density-matrix renormalization-group techniques and by comparing with the phase diagram of the classical-spin variant of the same model.
For a finite $t_{2}$, i.e., for a zigzag-ladder geometry, indirect antiferromagnetic interactions between the localized spins are geometrically frustrated. 
We demonstrate that $t_{2}$ at the same time {\em triggers} several magnetic phases which are absent in the model with nearest-neighbor hopping only. 
For strong $J$, we find a transition from antiferromagnetic to incommensurate magnetic short-range order, which can be understood entirely in the classical-spin picture. 
For weaker $J$, a spin-dimerized phase emerges, which spontaneously breaks the discrete translation symmetry.
The phase is not accessible to perturbative means but is explained, on a qualitative level, by the classical-spin model as well. 
Spin dimerization alleviates magnetic frustration and is interpreted as a key to understand the emergence of quasi-long-range spiral magnetic order which is found at weaker couplings.
The phase diagram at weak $J$, with gapless quasi-long-range order on top of the two-fold degenerate spin-dimerized ground state, competing with a nondegenerate phase with gapped spin (and charge) excitations, is unconventional and eludes an effective low-energy spin-only theory.
\end{abstract} 

\maketitle 

\section{Introduction}
\label{sec:intro}

One-dimensional lattice models have served as important paradigms for unconventional states of matter. 
This holds for pure spin models, such as the spin-$S$ Heisenberg model, and for interacting models of itinerant electrons, such as the Hubbard model, for example. 
The one-dimensional Kondo lattice with nearest-neighbor hopping $-t_{1}$ ($t_{1}>0$) and local antiferromagnetic exchange coupling $J$ can be seen as a hybrid model including localized spins as well as itinerant electrons. \cite{TSU97b}
The model is interesting as it incorporates the competition between Kondo-singlet formation \cite{Hew93} and, on the other hand, the emergence of nonlocal magnetic correlations due to indirect magnetic couplings between the local spins mediated by the conduction-electron system. \cite{RK54,Kas56,Yos57}
This competition has been pointed out early. \cite{Don77} 
At half-filling and as function of $J$, however, there is no quantum-phase transition. \cite{SU99} 
Furthermore, the half-filled Kondo model for arbitrary $J$ and on an arbitrary but bipartite lattice is well known to have a unique total spin-singlet ground state. \cite{She96} 

\begin{figure}[b]
\includegraphics[width=0.9\columnwidth]{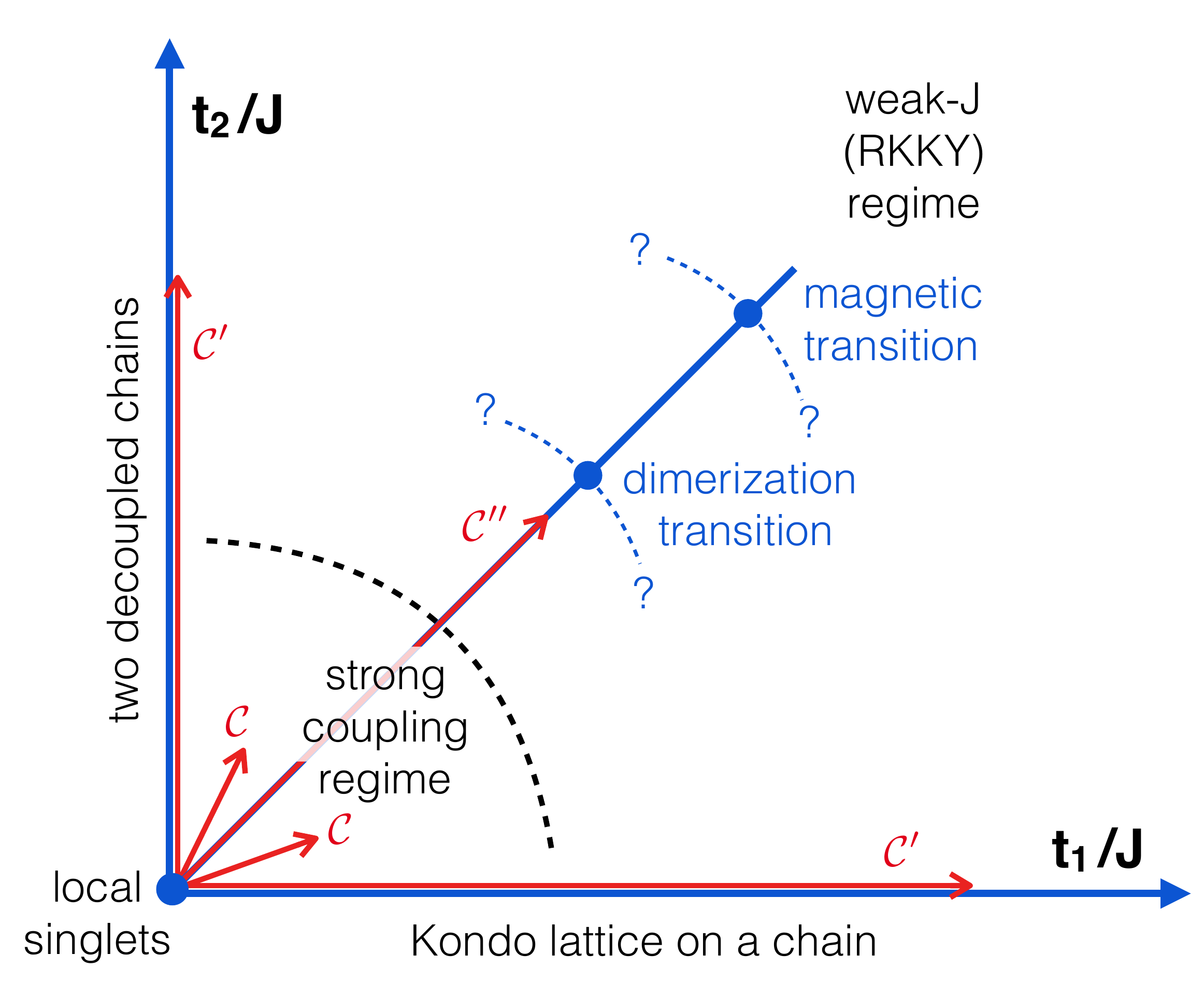}
\caption{
Sketch of the ground-state phase diagram of the Kondo lattice on the zigzag ladder.
Paths $\ca C$ adiabatically connect to the atomic limit $t_{1}=t_{2}=0$ (``local singlets'') via strong-coupling perturbation theory. 
The ground state of the unfrustrated Kondo lattice (on a chain with nearest-neighbor hopping only) or on two decoupled chains, adiabatically connects to the atomic limit for all $J>0$, as indicated by paths $\ca C'$. 
Along $\ca C''$ ($t_{1}=t_{2}$) a spin-dimerization and a magnetic phase transition have been found in Ref.\ \onlinecite{PRP18}.
}
\label{fig:toverj}
\end{figure}

Much less is known for the non-bipartite model with hopping beyond nearest neighbors $-t_{2}$.
The one-dimensional Kondo lattice with nearest-neighbor hopping $-t_{1}$ and next-nearest-neighbor hopping $-t_{2}$, is equivalent with the Kondo model on a zigzag ladder with hopping along the rungs $-t_{1}$ and along the legs $-t_{2}$. 
In the strong-coupling limit, i.e., for all $0 \le t_{1},t_{2} \ll J$, the state of the system is a featureless Kondo insulator and is obtained by nondegenerate perturbation theory from the ground state of the atomic-limit  $t_{1} = t_{2} = 0$, given by a simple tensor product of completely local Kondo-singlet states.
This adiabatic connection is indicated by the paths $\ca C$ in Fig.\ \ref{fig:toverj}. 
Actually, see paths $\ca C'$ in the figure, the adiabatic connection to the atomic limit not only holds for strong $J$ but for all finite $J$ in two limits, the single-chain limit $t_{1} \ne 0$, $t_{2}=0$, and the limit of two decoupled chains $t_{1} = 0$ but $t_{2}\ne 0$.
Here, previous exact-diagonalization calculations and density-matrix renormalization-group (DMRG) studies \cite{SU99} have demonstrated that the system is an insulator with gapped spin and charge excitations and exponentially decaying two-point correlations for all $J$.
Only in the (nonperturbative) limit $J \to 0$, the spin gap is expected to get exponentially small. \cite{TSU97b,SU99} 

Apart from the more or less trivial limits of a single chain or of two decoupled chains, the zigzag Kondo ladder at half-filling has only been studied recently, namely for $t_{1}=t_{2}$, see Ref.\ \onlinecite{PRP18} and the path $\ca C''$ in Fig.\ \ref{fig:toverj}.
Our DMRG study has in fact uncovered the presence of (at least) two quantum-phase transitions along $\ca C''$. 
Starting from the strong-$J$ limit, the system first undergoes a continuous (or at most a weakly first-order) transition at $\jd = \jdv$ to a spin-dimerized state. 
This is indicated by a finite dimerization order parameter 
$O_{D} = | \langle \ff S_{i-1} \ff S_{i} \rangle - \langle \ff S_{i} \ff S_{i+1} \rangle|$. 
Second, at $\jm = \jmv < \jd$, the system develops quasi-long-range spiral magnetic order. 
With decreasing $J$, the spin gap closes at $\jm$, and the spin-structure factor diverges at the wave vector $Q=\pi /2$, while charge excitations remain gapped at any coupling strength.

The emergence of magnetic order in this model is highly unconventional for several reasons: 
It only shows up in the {\em magnetically frustrated} system while the unfrustrated model with $t_{1}=0$ or with $t_{2}=0$ is paramagnetic. 
Furthermore, the magnetic ordering also eludes a simple perturbative explanation. 
The relevant parameter regime is well beyond the weak-coupling (Ruderman-Kittel-Kasuya-Yoshida, RKKY) regime \cite{RK54,Kas56,Yos57} and also beyond the strong-coupling regime, where a superexchange-like perturbation theory applies.
The proximity to the spin-dimerization transition, however, provokes an idea that has also been suggested for the two-dimensional Kondo lattice \cite{MNYU10,SAG18} and that has been tested on the dynamical \cite{AAP15} and static mean-field level, \cite{HUM11} namely that a spontaneous breaking of spatial symmetries, alleviating the magnetic frustration, also paves the way for a subsequent magnetic transition, and should thus be seen as a precursor of magnetism. 
Finally, the spiral nature of the magnetic state with the spin-structure factor diverging at $Q=\pi/2$ is reminiscent of the magnetism of {\em classical} spin degrees of freedom and of the generic compromise in the presence of geometrical frustration in classical spin models.

The purpose of the present study is to systematically trace the spin-dimerization and the magnetic phase transition in the $t_{1}$-$t_{2}$-$J$ parameter space and to check whether one can connect to a perturbatively accessible parameter regime. 
To this end, we employ a recently developed \cite{RP16} density-matrix renormalization group (DMRG) \cite{Sch11} code.
We furthermore map out the complete phase diagram of the model with quantum spins replaced by classical spins. 
This allows us to check to what degree quantum fluctuations are essential in explaining the phase diagram and provides a different and independent view on the problem.
While the classical, or actually quantum-classical hybrid model is expected to have a magnetically long-range-ordered ground state, the presence or absence of a dimerized state is of particular interest.

The paper is organized as follows: 
The model and the notations are introduced in the next section \ref{sec:mod}, and some issues of the numerical methods are discussed in Sec.\ \ref{sec:method}. 
Results for the classical-spin case are presented in Sec.\ \ref{sec:resclass} and compared to perturbation theory in Sec.\ \ref{sec:pert}.
The DMRG results are presented in Sec.\ \ref{sec:res}, and an extended discussion and our conclusions are given in Sec.\ \ref{sec:dis}.

\section{Frustrated Kondo lattice model}
\label{sec:mod}

The Hamiltonian of the Kondo-lattice model is given by 
\be
H = \sum_{ij \sigma} t_{ij} c^{\dagger}_{i\sigma} c_{j\sigma} 
+ 
J \sum_{i} \ff s_{i} \ff S_{i}
\: .
\label{eq:ham}
\ee 
Here, $c^{\dagger}_{i\sigma}$ ($c_{i\sigma}$) creates (annihilates) an electron at site $i=1, ..., L$ with spin projection $\sigma=\uparrow, \downarrow$, and $\ff s_{i} = \frac{1}{2} \sum_{\sigma \sigma'} c^{\dagger}_{i\sigma} \ff \tau_{\sigma\sigma'} c_{i\sigma'}$ is the local conduction-electron spin at site $i$, where $\ff \tau$ denotes the vector of Pauli matrices.
The local spin $\ff s_{i}$ couples antiferromagnetically, with interaction strength $J>0$, to the localized spin $\ff S_{i}$ at the same site.
Two model variants will be studied: 
(i) the conventional Kondo lattice where the latter is taken as a quantum spin with $S=1/2$, and
(ii) the corresponding classical-spin variant where $\ff S_{i}$ is treated as a classical vector of fixed length $|\ff S_{i}|=1/2$.
Furthermore, we consider the system at half-filling with $N=L$ electrons, where $L$ is the number of lattice sites.

\begin{figure}
\includegraphics[width=0.8\columnwidth]{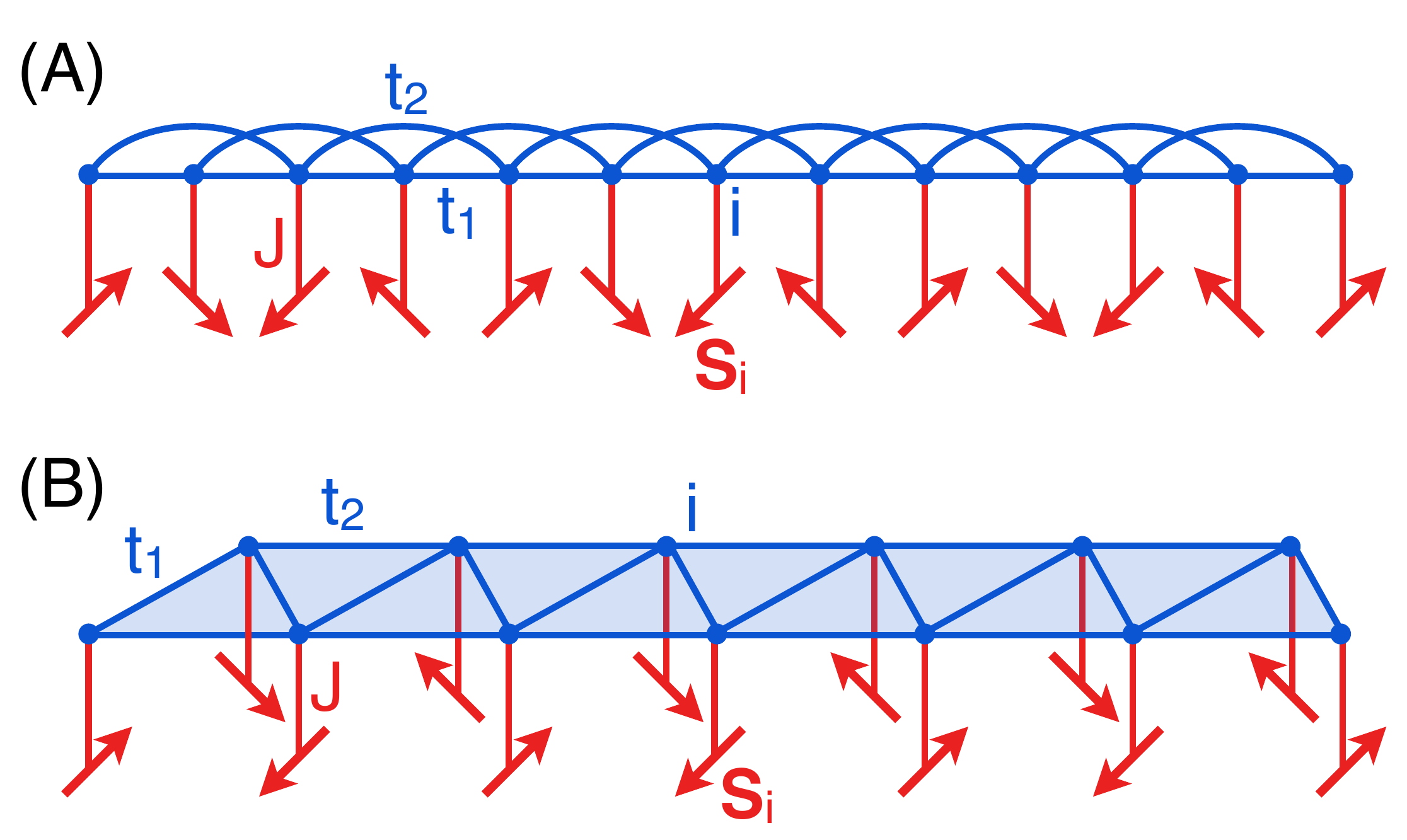}
\caption{
(A) One-dimensional Kondo-lattice model with nearest-neighbor hopping $-t_{1}$ and next-nearest-neighbor hopping $-t_{2}$, parameterized as $t_{1} = t \cos \varphi$ and $t_{2} = t \sin \varphi$ with $t>0$ and $0\le \varphi \le \pi/2$.
Local antiferromagnetic exchange $J>0$, see Eq.\ (\ref{eq:ham}).
(B) Equivalent representation of the same model. 
On the zigzag ladder, sites linked by $t_{1}$ on the rungs of the ladder or by $t_{2}$ along the legs of the ladder will both be denoted as nearest neighbors.
For $\varphi = 0$, i.e., $t_{1}=t$ and $t_{2}=0$, the model reduces to the Kondo lattice on the one-dimensional chain. 
At $\varphi=\pi/4$, we have $t_{1}=t_{2}$, and for $\varphi = \pi/2$, i.e., $t_{1}=0$ and $t_{2}=t$, the model is given by two decoupled one-dimensional chains. 
}
\label{fig:geo}
\end{figure}

The first term in Eq.\ (\ref{eq:ham}) describes the hopping of the conduction electrons on a one-dimensional lattice with hopping amplitude $t_{ij} = -t_{1}$ between nearest neighbors $i,j$ and with hopping $-t_{2}$ between next-nearest-neighbors.
We parameterize the hopping as $t_{1} = t\cos\varphi$ and $t_{2}=t \sin \varphi$ where $0\le \varphi \le \pi /2$ and $t>0$.
Fig.\ \ref{fig:geo} (A) provides a sketch of the geometry. 
Nearest neighbors are linked by the hopping $t_{1}$.
An equivalent view is the zigzag-ladder geometry (B). 
For convenience, we refer to nearest neighbors for both, a pair of sites linked by $t_{1}$, i.e., on the rungs of the ladder, and a pair of sites linked by $t_{2}$, i.e., along the legs of the ladder.
The model studied in Ref.\ \onlinecite{PRP18} is recovered for $\varphi=\pi/4$ ($t_{1}=t_{2}$).
Whenever convenient, we set $t \equiv 1$ to fix the energy unit.

\section{Methods}
\label{sec:method}

\subsection{Matrix-product states}
\label{sec:mps}

For the numerical solution of the Kondo lattice model, Eq.\ (\ref{eq:ham}), we employ a conventional density-matrix renormalization-group (DMRG) algorithm \cite{RP16,Sch11} as well as a recently suggested variationally uniform matrix-product state approach (VUMPS). \cite{ZSVF+18}
The DMRG calculations are based on a single-site algorithm, where we make use of a subspace expansion scheme as recently suggested in Ref. \onlinecite{HMSW15}. 
Our implementation explicitly takes into account the continuous symmetries of the Kondo lattice, namely the invariance under U(1) gauge transformations related to conservation of the total particle number as well as the SU(2) spin-rotation symmetry related to conservation of the total spin.
Here, we follow previous work detailed in Refs.\ \onlinecite{McC07,Wei12}. 
Exploiting the SU(2) symmetry is particularly important for the present case and leads to a substantial gain in efficiency and accuracy. 
The VUMPS calculations follow the recently developed method for obtaining the ground state directly in the thermodynamic limit. \cite{ZSVF+18}
The method explicitly makes use of the translational symmetry of the system to get the exact matrix-product-state representation of the ground state for one-dimensional gapped Hamiltonians. 
For gapless systems, the method yields an approximation to the ground state only, and all observables need to be scaled to the infinite-bond-dimension limit. 
Our VUMPS implementation respects the U(1) and SU(2) gauge symmetries.

For the DMRG calculations presented here, we keep up to $m \approx 8,000$ density-matrix eigenstates.
These are grouped into symmetry blocks labeled with the irreducible representations $(N,S)$ of U(1) and SU(2), respectively. 
This would correspond to $m_{\rm tot} \approx 40,000$ states if only Abelian symmetries were used. 
For the VUMPS calculations we keep up to $m \approx 15,000$ states, which corresponds to  $m_{\rm tot} \approx 100,000$ states in the Abelian case.

\subsection{Classical spins}

The DMRG results will be compared with those obtained for the classical-spin variant of the Kondo lattice. 
Generally, the ground-state spin configuration is obtained by minimization of the energy
\be
  E(\{\ff S\})
  =
  \sum_{ii'\sigma\sigma'} \left( t_{ii'} \delta_{\sigma\sigma'} + \frac{J}{2} (\ff\tau \ff S_{i})_{\sigma\sigma'} \delta_{ii'} \right)
  \langle c^{\dagger}_{i\sigma} c_{i'\sigma'} \rangle_{\{\ff S\}}
\: .
\label{eq:efunc}
\ee
as a functional of the spin configuration $\{\ff S\} = (\ff S_{1}, ..., \ff S_{L})$ subject to the constraints $|\ff S_{i}|=1/2$.
The hopping correlation, i.e., the elements 
\be
  \rho_{ii'\sigma\sigma'} 
  = 
  \langle c^{\dagger}_{i'\sigma'} c_{i\sigma} \rangle
\ee
of the $2L\times 2L$ one-particle reduced density matrix $\ff \rho$ are obtained via $\ff \rho = \Theta(- \ff t_{\rm eff}) = \ff U \Theta(-\ff \varepsilon) \ff U^{\dagger}$ from the effective Hermitian hopping matrix $\ff t_{\rm eff}$ with elements
\be
   t_{\rm eff,ii',\sigma\sigma'} = t_{ii'} \delta_{\sigma\sigma'} + \frac{J}{2} (\ff\tau \ff S_{i})_{\sigma\sigma'} \delta_{ii'}  
\label{eq:effhopp}
\ee
by diagonalization, $\ff t_{\rm eff} = \ff U \ff \varepsilon \ff U^{\dagger}$, where $\ff \varepsilon$ is the diagonal matrix of the eigenvalues of $\ff t_{\rm eff}$ for the given spin configuration and where $\Theta$ denotes the Heaviside step function.

\begin{figure}
\includegraphics[width=0.55\linewidth]{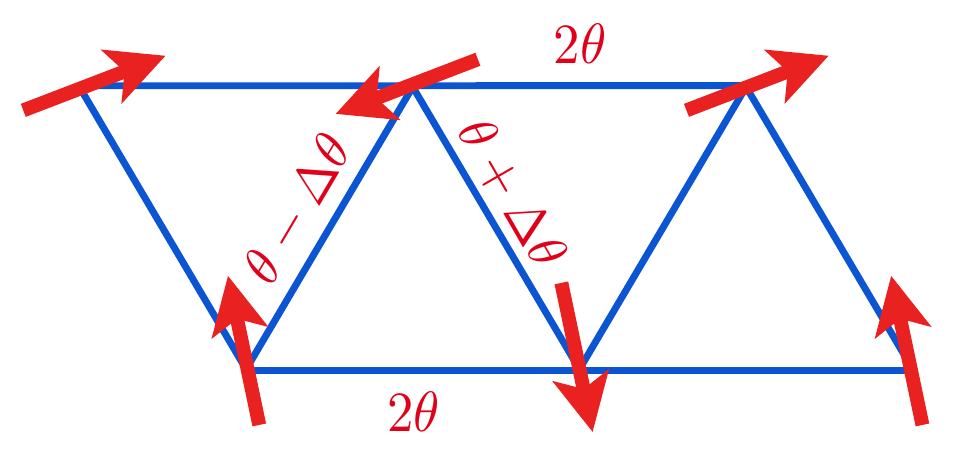}
\caption{
Parameterization of possible classical spin configurations.
$2 \theta$: angle between neighboring spins along the legs of the zigzag ladder. 
$\theta + \Delta \theta$ and $\theta - \Delta \theta$: alternating angles between neighboring spins along the rungs.
$\Delta \theta \ne 0$ indicates a dimerized state. 
Incommensurate spiral states with pitch angle $\theta$ are described with $\Delta \theta = 0$.
For the given example spin configuration, $\theta=\pi/2$, $\Delta \theta >0$.
It is sufficient to consider the parameter ranges $0\le \theta \le \pi$ and $0\le \Delta \theta \le \pi/2$. 
}
\label{fig:theta}
\end{figure}

Antiferro- and ferromagnetic spin configurations as well as spiral phases with arbitrary pitch angle $\theta$ and spin-dimerized phases can be covered with a $(\theta, \Delta \theta)$ parameterization of the classical spin configurations, see Fig.\ \ref{fig:theta}. 
Note that the transformation $\theta \to \pi - \theta$ and $\Delta \theta \to \pi - \Delta \theta$ is a symmetry. 
Assuming periodic boundary conditions, the allowed pitch angles are given by $\theta = n \, 2\pi / L$ with integer $n$. 
For convenience, the same grid is used for $\Delta \theta$.
Minimization of the corresponding energy function $E=E(\theta, \Delta \theta)$ is typically performed numerically for systems  with up to $L=200$ lattice sites. 
In the range $\pi / 4 \lesssim \varphi < \pi /2$ and for strong $J$, however, converged results can only be achieved for systems as large as $L = \ca O (10^{5})$ (see discussion in Sec.\ \ref{sec:pertstrong}). 

\section{Ground-state phase diagram for classical spins}
\label{sec:resclass}

\begin{figure}
\includegraphics[width=0.95\columnwidth]{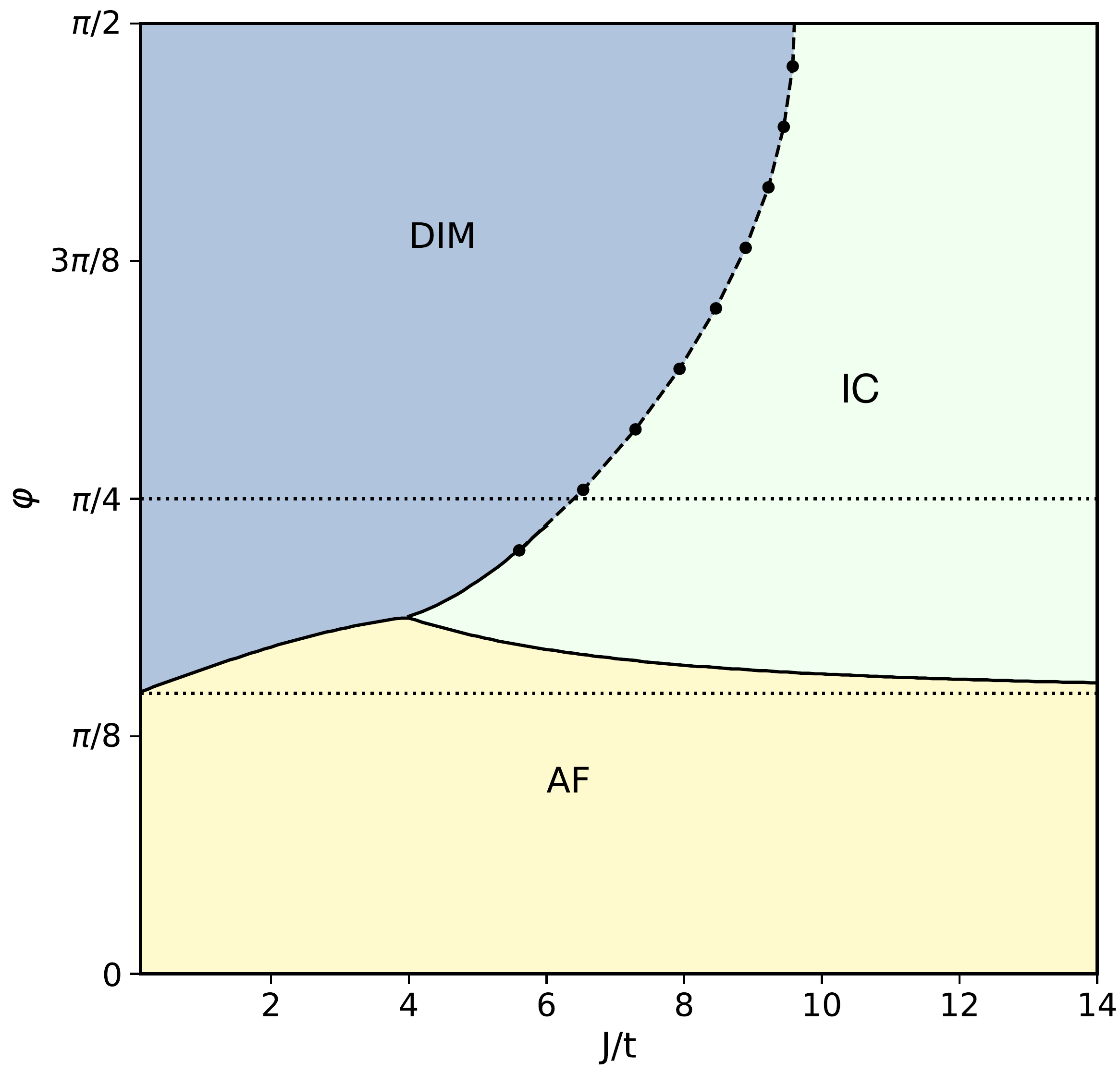}
\caption{
$\varphi$-$J$ magnetic phase diagram of the classical-spin variant of the Kondo lattice
with an antiferromagnetic phase (AF, $\theta = \pi$, $\Delta \theta=0$), an incommensurate spiral phase (IC, $\pi/2 < \theta < \pi$, $\Delta \theta=0$), and a dimerized phase (DIM, $\theta = \pi/2$, $\Delta \theta=\pi /2$).
The dashed line indicates $t_{1} = t_{2}$ ($\varphi = \pi / 4$).
Calculations have been performed for $L=200$. 
This is sufficient for convergence, except for the regime $\varphi \gtrsim \pi /4$ where much larger systems with up to $L = 100,000$ sites are necessary (see text for discussion). 
The dotted line interpolates between the data points.
}
\label{fig:pdclass}
\end{figure}

We start with the discussion of the classical-spin case.
The resulting phase diagram is shown in Fig.\ \ref{fig:pdclass} for the entire $\varphi$-range and for coupling strengths $J$ spanning the parameter range between the perturbatively accessible regimes of weak ($J\ll t$) and strong coupling $J \gg t$.
Due to the missing quantum fluctuations, the ground state exhibits long-range magnetic order and is degenerate with respect to global SO(3) rotations of the classical spins.
We expect, however, that the type of magnetic order is related to the corresponding type of magnetic short-range correlations or quasi-long-range (algebraic) magnetic order of the quantum-spin case since the classical-spin approach comprises the relevant indirect magnetic coupling mechanisms at work. 

For $\varphi=0$, i.e., for the unfrustrated chain, we find the expected antiferromagnetic order ($\theta=\pi$, $\Delta \theta = 0$) for all coupling strengths $0 < J < \infty$. 
This corresponds to the well-known quantum-singlet state with short-range antiferromagnetic correlations. \cite{TSU97b,SU99}
At weak $J$ this is driven by the antiferromagnetic RKKY indirect magnetic interaction between the classical spin, while for strong $J$ a superexchange-like mechanism is responsible (see below).

As the classical-spin system acts like a staggered magnetic field, the magnetic unit cell is given by twice the unit cell of the lattice, and the electronic band structure consists of two dispersive bands in the reduced Brillouin zone $-\pi/2 < k \le \pi /2$ with a gap of $JS=J/2$ at the zone boundary. 
Hence, the system is an insulator. 

Switching on the hopping $t_{2}$ on the rungs of the zigzag lattice, the system stays insulating. 
For all magnetic phases found, including the incommensurate spiral phase, there is a finite gap between the highest occupied and the lowest unoccupied one-particle energies, given by the eigenvalues of the effective hopping matrix, Eq.\ (\ref{eq:effhopp}). 
For very weak $J$, however, a somewhat more cautious statement is appropriate. 
As the gap is of the order of $J$, it becomes of the order of the finite-size gap $\sim t/L$ for $J\to 0$ at fixed $L$ such that no clear statement is possible for $J \lesssim t / L$.

\begin{figure}
\includegraphics[width=0.95\columnwidth]{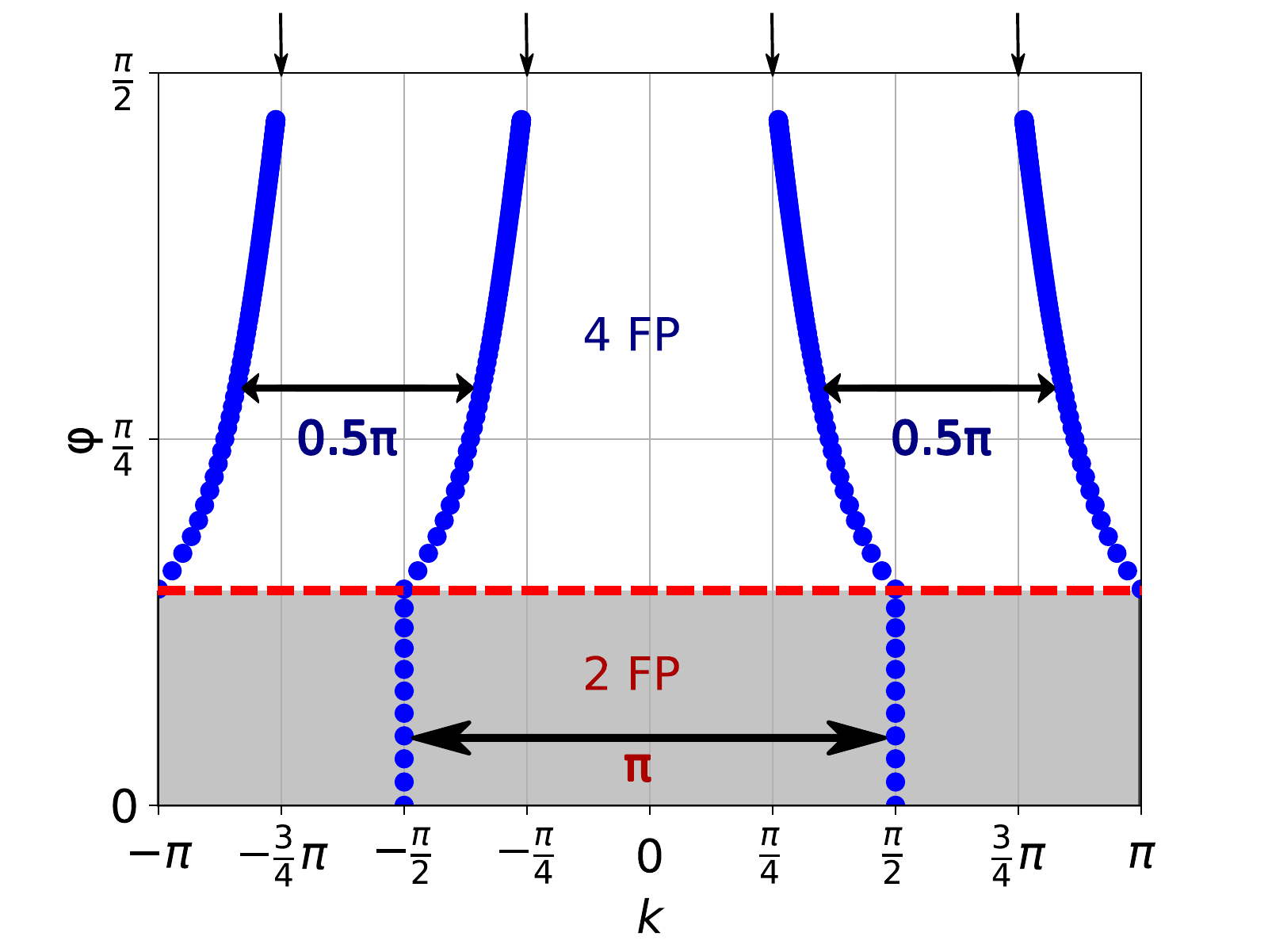}
\caption{
Location of the Fermi points in $k$-space as a function of $\varphi$.
The Lifshitz transition between a state with two (2 FP) and with four Fermi points (4 FP) takes place at $\varphi= \varphi_{0} = \arctan (\frac 12) \approx 0.148 \pi$, i.e., at $t_{1} = 2 t_{2}$ (dashed red line).
Horizontal arrows: nesting vectors.
Vertical arrows mark the decoupled-chains limit $\varphi = \pi /2$.
}
\label{fig:fermi}
\end{figure}

For finite $\varphi > 0$, the antiferromagnetic phase extends in the entire $J$ range up to some $J$-dependent critical value $\varphi_{\rm c}(J)$.
In the weak-coupling limit, we find $\varphi_{\rm c}(J\to 0) \equiv \varphi_{0} \approx 0.148 \pi$. 
This corresponds to $t_{2} = t_{1}/2$ or, equivalently, to $\varphi_{0} = \arctan(\frac 12)$, which is exactly the Lifschitz point where the number of Fermi points in the noninteracting band structure changes from two (for $\varphi < \varphi_0$) to four (for $\varphi > \varphi_0$), see Fig.\ \ref{fig:fermi}.
In the strong-$J$ limit, on the other hand, the critical value is
found as $\varphi_{\rm c}(J\to \infty) \equiv \varphi_{\infty} \approx
0.148\pi$.
This perfectly reproduces the result of strong-coupling perturbation theory, namely $\varphi_{\infty} = \arctan(\frac 12)$, see Eq.\ (\ref{eq:ptres2}) below, which is a coincidental match with the value for $\varphi_{0}$.

At $J = J_{\rm tri} \approx 4.1t$ there is a ``triple point'' on the critical line $\varphi_{c}(J)$ (with $\varphi_{\rm tri} = \varphi_{c}(J_{\rm tri}) \approx 0.188 \pi$). 
Above the line $\varphi_{\rm c}(J)$, we find a spin-dimerized phase with $\theta = \pi/2$ and $\Delta \theta =\pi /2$ for weaker $J$. 
Here, the spins are ferromagnetically aligned on every second rung and antiferromagnetically between the rungs. 
This phase is separated by a line $\varphi_{\rm c}^{\rm (dim)}(J)$ from an incommensurate spin-spiral phase for stronger $J$ with $\Delta \theta = 0$ and with a continuously varying pitch angle in the range $\pi / 2 < \theta <\pi$. 
$\varphi_{\rm c}^{\rm (dim)}(J)$ is a monotonously decreasing function with decreasing $J$ and, at $J=J_{\rm tri}$ terminates at the triple point. 

The transition between the antiferromagnetic (AF) and the dimerized phase (DIM), see Fig.\ \ref{fig:pdclass}, is discontinuous.
Across the line $\varphi_{\rm c}(J)$, there is a finite jump of the optimal values for $\theta$ and $\Delta \theta$ as obtained by minimization of the total-energy function $E(\theta,\Delta \theta)$, i.e., the local minimum at  
$(\theta = \pi,\Delta \theta=0)$ becomes degenerate with the local minimum at 
$(\theta = \pi/2,\Delta \theta=\pi/2)$. 
Similarly, transitions between the DIM and the incommensurate spiral phase (IC) are
discontinuous, opposed to transitions between the AF and IC phases which turn out as continuous. 
In the latter case, and as a function of $\varphi$, for example, the total-energy minimum at $(\theta = \pi,\Delta \theta=0)$ continuously shifts from $\theta = \pi$ to $\theta < \pi$, or vice versa, while the derivative $\partial E / \partial \varphi$ shows a finite jump across the critical line.

The most remarkable result of classical-spin theory consists in the prediction of the spin-dimerized phase. 
This can be seen as a simple effective theory of the spin dimerization found in the full quantum-spin Kondo lattice. \cite{PRP18}
Clearly, due to the mean-field character inherent to the classical-spin theory, one cannot expect correct order of magnitude for the critical parameters but still the comparison with the exact DMRG results is instructive. 
One would expect that the necessary critical interaction strength $J_{\rm c,class}^{\rm (dim)}$ would be much stronger than $\jd$, the DMRG value, since mean-field-like and classical approaches tend to overestimate ordering, i.e., dimerization due to absence of quantum fluctuations that act against ordering.
Indeed, on the $t_{1}=t_{2}$ line, for example, we find $6.4t \approx J_{\rm c,class}^{\rm (dim)} \gg \jd \approx \jdva$ instead, see Ref.\ \onlinecite{PRP18}. 
Below we will argue that spin dimerization cannot occur in the strong-$J$ limit and it is rather the alleviation of frustration which drives spin dimerization.

\section{Perturbative approaches}
\label{sec:pert}

\subsection{Strong Kondo coupling $J$}
\label{sec:pertstrong}

In the ground state of the atomic limit $t_{1}=t_{2}=0$, each lattice site $i$ is occupied by exactly one electron. 
The local spin moment $\langle \ff s_{i} \rangle$ is fully polarized and oriented antiparallel to the classical spin $\ff S_{i}$. 
To compute the functional $E(\{\ff S\})$, see Eq.\ (\ref{eq:efunc}), the configuration of classical spins $\{\ff S\} = (\ff S_{1}, ..., \ff S_{L})$ must be considered as fixed, and thus the electronic ground state $|0\rangle$ is nondegenerate. 
The first nonzero contribution to the functional within nondegenerate perturbation theory in powers of $t/J$ is found at second order: 
\be
  E(\{\ff S\}) = - L \frac{J}{4} + \sum_{n \ne 0} \frac{|\langle 0 | H_{1} | n \rangle|^{2}}{E_{0}-E_{n}} + \ca O(t^{3}/J^{2}) \: .
\label{eq:eofspt}  
\ee
Here, the perturbation $H_1 = \sum_{ij\sigma} t_{ij} c^\dagger_{i\sigma} c_{j\sigma}$ is the hopping term in Eq.\ (\ref{eq:ham}), $| 0 \rangle$ and $|n \rangle$ are the ground and the excited states of the Kondo term $H_{0}$, and $E_{0} = - LJ/4$ and $E_{n}$ are the corresponding unperturbed eigenenergies. 
The straightforward calculation is given in the Appendix \ref{sec:pt} and results in: 
\begin{eqnarray}
  E(\theta,\Delta\theta) 
  &=& \mbox{const.} + \frac L4 
  \left(  
  J_1 \cos \theta \cos \Delta \theta  + J_2 \cos (2\theta )
  \right)
  \nonumber \\
  &+& \ca O(t^{4}/J^{3})  \: ,
\label{eq:ptres1}  
\end{eqnarray}
which is just the energy of the classical-spin ($|\ff S_{i}| = 1/2$) Heisenberg model on the zigzag ladder with exchange couplings $J_1 = 8t_1^2/J$ and $J_2 = 8t_2^2 / J$ and for the same parameterization of the spin configuration as assumed above for the classical-spin Kondo lattice (see Appendix \ref{sec:heis}).
The constant, $\{\ff S\}$-independent energy offset is given by Eq.\ (\ref{eq:offset}) of the Appendix \ref{sec:pt}.

We note that Eq.\ (\ref{eq:ptres1}) holds up to fourth-order corrections. 
Namely, the ground-state energy correction at order $t^{3}/J^{2}$ vanishes identically for all $(\theta, \Delta \theta)$ due to a cancellation of two different types of virtual ring-exchange processes. 
The according calculation is a bit more tedious but still straightforward and is not reported here.

Minimization of the energy functional yields $\Delta \theta = 0$, i.e., there is no spin dimerization in this limit. 
We furthermore get $\theta = \pi$ for $t_1 > 2 t_2$ and 
\be
  \theta = \arccos\left(-\frac{t_1^2}{4t_2^2}\right) = \arccos \left( -\frac{1}{4\tan[2](\varphi)} \right)
\label{eq:ptres2}  
\ee
for $t_1 < 2 t_2$.

Strong-coupling perturbation theory also explains why convergence of the results with increasing $L$ is extremely poor in the range $\pi / 4 \lesssim \varphi < \pi /2$ and for strong $J$. 
As detailed in Appendix \ref{sec:fs}, comparatively large systems must be considered to control the finite-size effects. 
Calculations in this parameter regime are performed for systems with up to $L = 100,000$ sites, see dots in Fig.\ \ref{fig:pdclass}.  

\subsection{RKKY theory at weak $J$}

Standard RKKY theory \cite{RK54,Kas56,Yos57} provides us with an effective spin Hamiltonian 
\be
  H_{\rm RKKY} = \sum_{k} J_{\rm RKKY}(k) \ff S_{k} \ff S_{-k}
\label{eq:rkkyham}  
\ee
at order $J^{2}$ in the limit $J \to 0$. 
The effective RKKY coupling $J_{\rm RKKY}(k) = - J^{2} \chi_{0}(\omega=0,k)$ is given by the static magnetic susceptibility of the conduction electrons
\be
  \chi_{0} (\omega=0,k) 
  = 
  \frac{1}{2L} \sum_{q} \frac{n_{k+q,\uparrow} - n_{q,\downarrow}}
  {\varepsilon(q) - \varepsilon(k+q)} \: ,
  \label{eq:chi0}
\ee
where $n_{k\sigma} = \Theta(-\varepsilon(k))$ is the occupation number.
For the zigzag lattice, the dispersion reads
\be
  \varepsilon(k) = -2t_{1} \cos(k) -2t_{2} \cos(2k) \: . 
\ee

If $t_{2} < t_{1} / 2$ ($\varphi < \varphi_0 = \arctan(\frac12) \approx 0.148\pi$), there are two Fermi points at $k_{\rm F} = \pm \pi / 2$, independent of $\varphi$ as dictated by Luttinger's sum rule. \cite{LW60,Lut60}
The susceptibility diverges logarithmically with $L\to \infty$ at $k=\pi$, as is easily seen by expanding the denominator in $q$ around $q=k_{\rm F}$.
Hence, spin correlations are predominantly antiferromagnetic, consistent with the antiferromagnetic phase found numerically.

If $t_{2}>t_{1}/2$, there are four Fermi points, see Fig.\
\ref{fig:fermi}, resulting in a logarithmic divergence of the
susceptibility $\chi_{0}(0,k)$ at the $\varphi$-independent nesting ``vector'' $k=\pi/2$.
This is consistent with a $\pi / 2$ spin spiral as well as with a spin-dimerized phase. 

We conclude that weak-coupling perturbation theory appears to explain the presence of the phase transition at $t_{1} = 2 t_{2}$ in the $J\to 0$ limit. 
One has to be aware, however, that the effective RKKY model (\ref{eq:rkkyham}) is actually ill-defined in one dimension \cite{TSU97b} due to the divergence of the coupling constant.

As a standard regularization let us consider arbitrary but {\em finite} $L$. 
Here weak-coupling perturbation theory is well behaved, and one may compute the RKKY coupling constants numerically using Eq.\ (\ref{eq:rkkyham}) or in the real-space representation as given in the supplemental material of Ref.\ \onlinecite{SGP12b}, for instance.
At half-filling $N=L$, we have to choose the system size as $L=4n+2$ with integer $n$ if $t_{2}<t_{1}/2$ ($\varphi < \varphi_{0}$) to get a nondegenerate electronic ground state, and $L=4n$ for $t_{2} > t_{1}/2$ ($\varphi > \varphi_{0}$). 
In the latter case, occupied two-fold spin-degenerate one-particle states labeled by wave vectors $k$ come in pairs $\pm k$, except for $k=0$ and $k=\pi$.
Restricting ourselves to homogeneous, to commensurate or incommensurate spin-spiral, and to spin-dimerized states, the ground-state classical-spin configuration is obtained by minimization of the energy function
\be
  E_{\rm RKKY}(\theta, \Delta \theta) = E_{0} + \sum_{ij} J_{{\rm RKKY,} ij} \ff S_{i} \ff S_{j} \: .
\label{eq:erkky}
\ee
Here, the constant offset $E_{0}$ is given by the total ground-state energy of the half-filled conduction band at $J=0$, i.e., $E_{0} = \sum_{ij\sigma} t_{ij} \langle c_{i\sigma}^{\dagger} c_{j\sigma} \rangle$.
Eq.\ (\ref{eq:erkky}) is easily evaluated numerically, and we find a nontrivial $\theta$ and $\Delta \theta$ dependence of $E_{\rm RKKY}(\theta, \Delta \theta)$. 

RKKY theory does not recover the spin-dimerized phase:
Computing the respective ground-state energies of the RKKY-Hamiltonian (\ref{eq:rkkyham}) for $\varphi > \varphi_{0}$ and for arbitrary but finite $L=4n$, shows that states with $\Delta \theta = \pi / 2$ and states with $\Delta \theta = 0$ are degenerate.
Hence, higher-order-in-$J$ perturbation theory would have to be invoked to lift this degeneracy and to reproduce the spin-dimerized phase that is found numerically within the full theory.

The full semiclassical theory predicts a discontinuous transition at $\varphi_{c}(J)$ between an antiferromagnetic ($\theta=\pi$, $\Delta \theta=0$) and a dimerized state ($\theta=\pi/2$, $\Delta \theta = \pi /2$), which are both characterized as local energy minima. 
At weak $J$ and for any finite $L$, we find 
\be
  \varphi_{c}(J) = \varphi_{0} + \mbox{const} \times J^2 + \ca O(J^{4}) \: .
\label{eq:phic}
\ee
It is important to understand that the $J^{2}$ term is already {\em beyond} second-order perturbation theory:
Expanding the energy of both minima ($i=1,2$) in powers of $J$, we have: 
\be
E_{i}(J,\varphi) 
= 
E_{i,0}(\varphi)
+
J^{2} E_{i,1}(\varphi)
+
J^{4} E_{i,2}(\varphi)
+
\ca O (J^{6})
\: . 
\ee
For $J=0$, the total energy does not depend on the spin configuration such that trivially $E_{1,0}(\varphi) = E_{2,0}(\varphi)$. 
Therefore, the $J^{2}$-term of $\varphi_{\rm c}(J)$ is obtained from the condition 
$E_{1,1}(\varphi_{\rm c})+J^{2} E_{1,2}(\varphi_{\rm c}) = E_{2,1}(\varphi_{\rm c})+J^{2} E_{2,2}(\varphi_{\rm c})$, i.e., one has to go to $\ca O(J^{4})$.
The RKKY Hamiltonian, on the other hand, involves a single energy scale only and thus predicts $\varphi_{\rm c}(J) = \varphi_{0} = \mbox{const}$ (and this relates to the transition w.r.t.\ $\theta$ only).
We note that $\varphi_{0}$ itself is independent of $L$ as this is related to the Lifschitz transition which is $L$-independent in turn. 

\begin{figure}
\includegraphics[width=0.93\columnwidth]{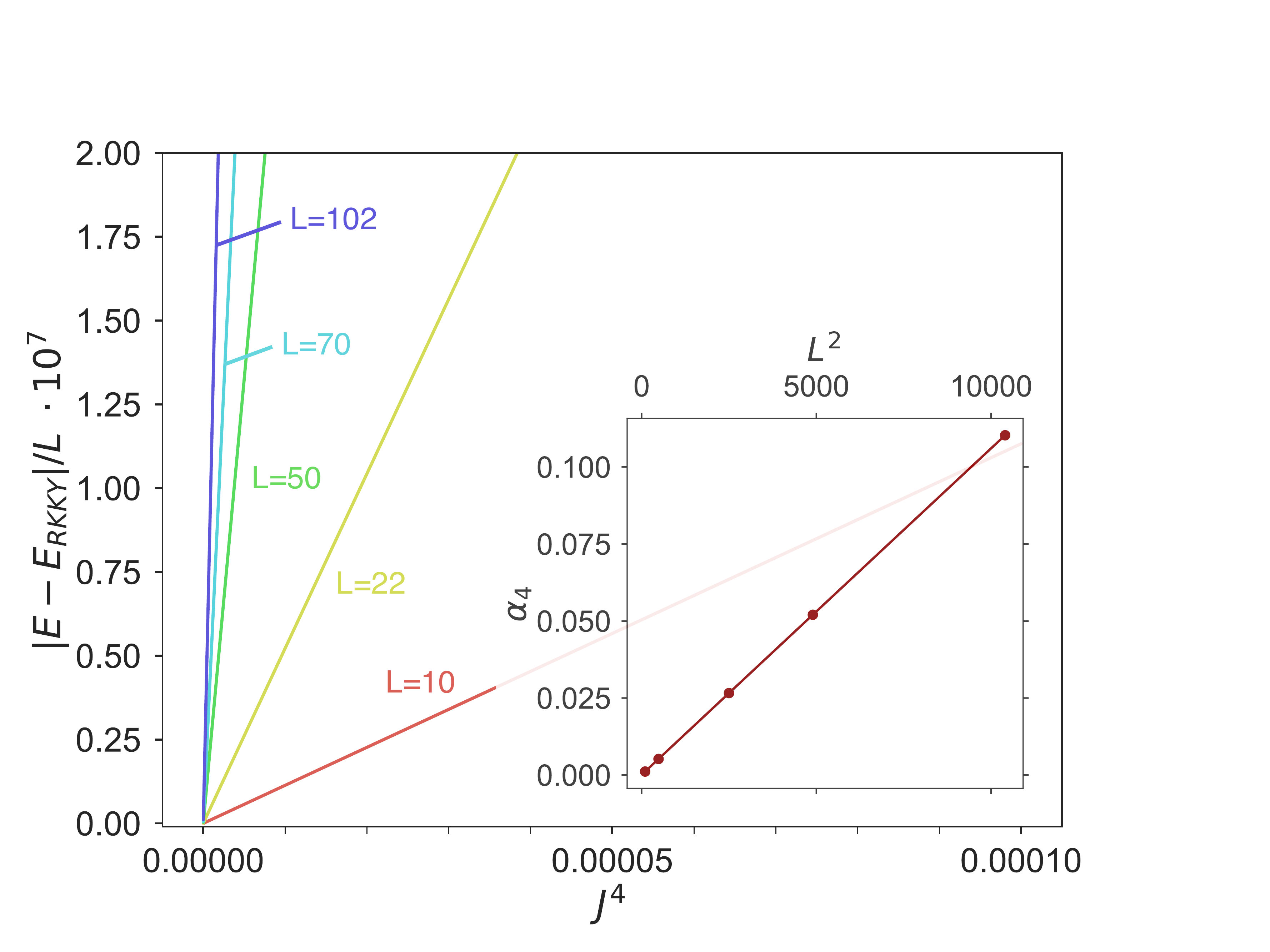}
\caption{
{\em Main figure:} Difference between the total energy of the full semiclassical theory and of the RKKY perturbation theory per site as a function of $J^4$ in the weak-$J$ regime at $\varphi=0.4 < \varphi_{0}$ and for different system sizes as indicated. 
{\em Inset:} Corresponding slope, i.e., the coefficient $\alpha_{4} \equiv \lim_{J \to 0} (E(J) - E_{\rm RKKY} ) / (L J^{4})$ as a function of $L^{2}$. 
}
\label{fig:rkky}
\end{figure}

Fig.\ \ref{fig:rkky} (main figure) demonstrates that RKKY theory is perfectly valid in the weak-$J$ limit for any {\em finite} system size $L$. 
Namely, the difference between the total energy, as obtained from the full semiclassical theory, and the RKKY energy, Eq.\ (\ref{eq:erkky}), is zero up to corrections of the order of $J^{4}$. 
We also note that for a small system with $L=10$ sites, for instance, there is almost perfect agreement between $E_{\rm RKKY}(\theta, \Delta \theta)$ and the exact ground-state energy $E(\theta,\Delta\theta)$ up to $J=0.5$ (not shown). 
In the same $J$ range but for larger lattices with $L=100$, however, there are qualitative deviations between the RKKY and the exact data for $E(\theta,\Delta\theta)$.
In fact, the magnitude of the $\ca O(J^{4})$ correction strongly increases with increasing system size, as can be seen in the main part of Fig.\ (\ref{fig:rkky}). 
This can be quantified by the coefficient $\alpha_{4} \equiv
E_{1,2}(\varphi) / L = \lim_{J \to 0} (E(J) - E_{\rm RKKY} ) / (L J^{4})$, i.e., by the slope of the linear trend with $J^{4}$. 
The inset demonstrates that $\alpha_{4}$ diverges as $L^{2}$ when $L\to \infty$.
This illustrates the breakdown of perturbation theory in the thermodynamical limit.

At the order $J^{2}$, on the other hand, $\alpha_{2} \equiv E_{1,1}(\varphi) =  \lim_{J \to 0} (E_{\rm RKKY} - E_{0} ) /(LJ^{2})$,
converges to a finite value as $L\to \infty$, as is well known for the
free electron gas. \cite{GVD05,Yaf87}
The logarithmic divergence of $\chi_{0}(0,k)$ at $k=\pi / 2$ is integrable, which implies that the total RKKY energy per site, Eq.\ (\ref{eq:rkkyham}), converges to a finite value in the thermodynamic limit. \cite{chi0}
We conclude that the $J^{2}$-term of $\varphi_{\rm c}(J)$, see Eq.\ (\ref{eq:phic}), is ill-defined when $L\to \infty$ since it is fixed by a condition involving both $\alpha_{2}$ and $\alpha_{4}$.

\subsection{Perturbation theory around $t_{1}=0$}

Finally, we briefly discuss perturbation theory in powers of $t_{1}$ around $t_{1}=0$, i.e., $\varphi=\pi/2$.
At $t_{1}=0$, the ground state is given by two decoupled antiferromagnetically ordered chains, where both, the dimerized and the incommensurate spiral configurations are degenerate. 
This is lifted at finite $t_{1}$ and produces a line $J_{\rm c}^{\rm (dim)}(t_{1})$ of first-order transitions. 
This can also be written as $J_{\rm c}^{\rm (dim)}(\varphi)$ or, in the notation used above, as $\varphi_{\rm c}^{\rm (dim)}(J)$. 
The phase diagram, Fig.\ \ref{fig:pdclass}, shows that for $t_{1} \to 0$, the line $J_{\rm c}^{\rm (dim)}(t_{1}) \to J_{0} \approx 9.7t$.
 
One may use nondegenerate perturbation theory in $t_{1}$ to compute the total energy for a given configuration of the classical spins. 
Dimerized states ($i=1$) with $\Delta \theta = \pi/2$ and spiral states ($i=2$) with $\Delta \theta = 0$ are given by local minima. 
We expand the energy of both minima ($i=1,2$) in powers of $t_{1}$,
\be
E_{i}(J,t_{1}) 
= 
E_{i,0}(J)
+
t_{1}^{2} E_{i,1}(J)
+
t_{1}^{4} E_{i,2}(J)
+
...
\: , 
\ee
and exploit the degeneracy $E_{1,0}(J) = E_{2,0}(J)$ in the decoupled-chain limit, where the energy is trivially independent of $\Delta\theta$. 
Note that odd powers of $t_{1}$ do not contribute.
The explicit computation at order $t_{1}^{2}$ is already somewhat tedious (and is not reported here), but it does not suffer from divergencies (as for perturbation theory in $J$), and yields the simple result $E_{1,1}(J) = E_{2,1}(J)$. 
This implies that the degeneracy is lifted, at the earliest, at order $t_{1}^{4}$. 
The condition fixing $J_{\rm c}^{\rm (dim)}(t_{1})$ then reads as: 
$E_{1,2}(J^{\rm (dim)}_{\rm c})+t_{1}^{2} E_{1,3}(J^{\rm (dim)}_{\rm c}) = E_{2,2}(J^{\rm (dim)}_{\rm c})+t_{1}^{2} E_{2,3}(J^{\rm (dim)}_{\rm c})$. 
We conclude that for an analytical computation of $J_{0}$ one would have to go $\ca O(t_{1}^{6})$ at least.
In any case we have $J_{c}^{\rm (dim)}(t_{1}) = J_{0} + \mbox{const.} \times t_{1}^{2} + \ca O(t_{1}^{4})$, or put differently, 
$J_{c}^{\rm (dim)}(\varphi) - J_{0} \propto (\varphi - \pi / 2)^{2}$.
Note that this is fully consistent with the dotted line in Fig.\ \ref{fig:pdclass} interpolating between the data points.

\section{DMRG results}
\label{sec:res}

\begin{figure}
\includegraphics[width=0.93\columnwidth]{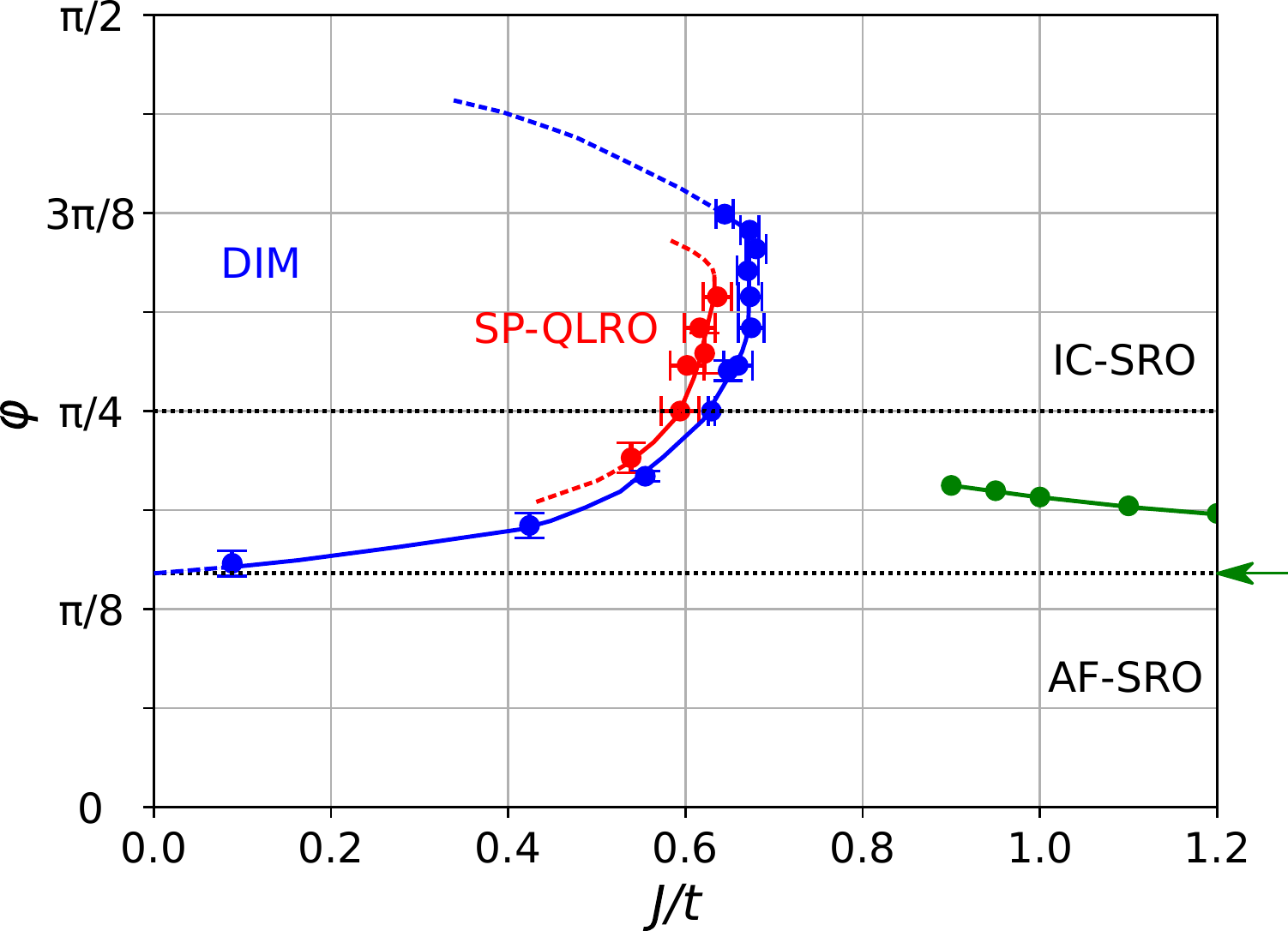}
\caption{
Ground-state phase diagram in the $J$-$\varphi$ plane for the half-filled quantum-spin Kondo model on the zigzag ladder, as obtained by DMRG calculations for systems of up $L=52$ sites (and extrapolated to $L=\infty$) as well as by VUMPS calculations 
working directly in the thermodynamical limit (with bond dimensions extrapolated to $m=\infty$ where necessary).
AF-SRO: antiferromagnetic short-range order with wave vector $Q=\pi$. 
IC-SRO: incommensurate spiral short-range order with $\pi /2 \le Q < \pi$.
DIM: spin-dimerized phase.
SP-QLRO: spiral quasi-long-range order, $Q=\pi /2$.
Points with error bars locate the various transitions. 
Black dashes lines: $t_2=t_1$ (upper) and $t_2=t_1/2$ (lower).
Blue and red dashed lines: see text. 
}
\label{fig:dmrg-phase}
\end{figure}

Let us now turn to the Kondo lattice with quantum spins $S=1/2$. 
The ground-state $\varphi$-vs.-$J$ phase diagram as obtained by extensive DMRG calculations is shown in Fig.\ \ref{fig:dmrg-phase}.
We will first give a rough overview over the different phases and parameter regimes and then proceed with a detailed discussion and the comparison with the classical-spin phase diagram.

For strong $J$, there are two different homogeneous phases, one with short-range antiferromagnetic spin correlations characterized by the wave-vector $Q=\pi$ (AF-SRO), and another one with short-range incommensurate (spiral) spin correlations (IC-SRO) characterized by a wave-vector in the range $\frac{\pi}{2} \le Q < \pi$, at which the spin-structure factor is at its maximum. 
Both phases are separated by the green line in Fig.\ \ref{fig:dmrg-phase}. 
The green arrow indicates the boundary in the $J\to \infty$ limit.

For weaker interaction strength, $J \lesssim 0.9t$, the spin-structure factor gets more complicated such that the ``phase boundary'' is no longer well defined.
In particular, with decreasing $J$, a second peak grows near $\frac{\pi}{2}$. 
This is a precursor of a gapless ground state with quasi-long-range $90^{\circ}$ spiral magnetic order (SP-QLRO, $Q=\frac{\pi}{2}$) which is found at still weaker $J$ (see red line).

The transition to this magnetic state, however, is preempted by a spin-dimerized phase (DIM) with spontaneously broken translation symmetry (blue line).
This is characterized by alternating ferro- and antiferromagnetic nearest-neighbor correlations on the rungs of the ladder.
In the weak-coupling limit the phase transition between the dimerized and the short-range antiferromagnetic states takes place at $\varphi_{\rm c}(J \to 0) = \arctan(\frac{1}{2}) \approx 0.148 \pi$ and exactly coincides with the transition point in the strong-coupling limit $\varphi_{\rm c}(J\to \infty)$ as is indicated by the dotted line.
Finally, the system is insulating in the whole phase diagram.
Charge excitations are gapped and the momentum-distribution function is does not show a singularity.

\subsection{Strong-$J$ regime}

\begin{figure}
\includegraphics[width=0.92\columnwidth]{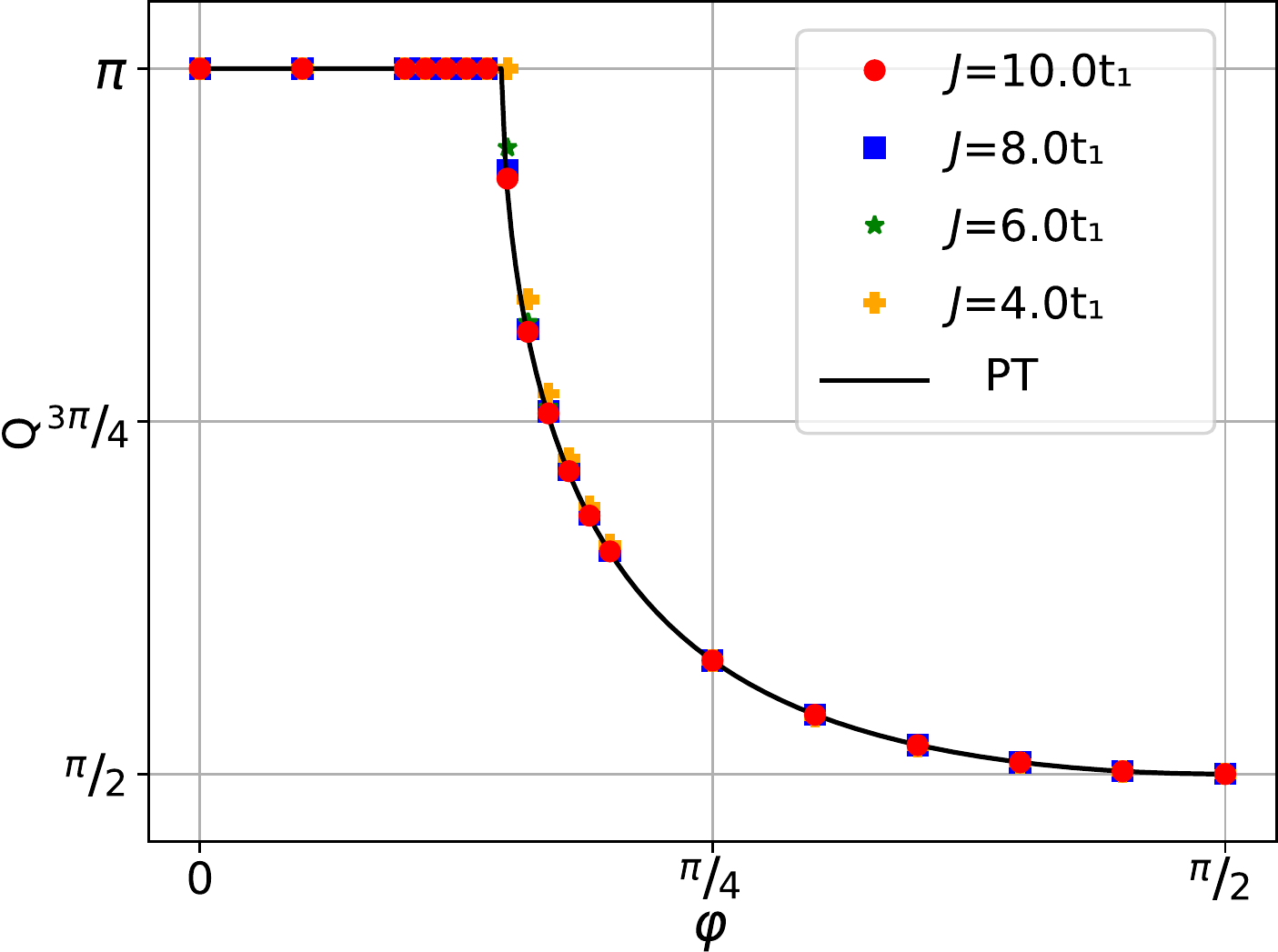}
\caption{
Position in the reciprocal unit cell $Q$ of the peak maximum in the spin-structure factor $S(Q)$, see Eq.\ (\ref{eq:ssf}), as a function of $\varphi$. 
The kink defines $\varphi_{\rm c}(J)$, i.e., the crossover from short-range AF to IC magnetic order.
Results of VUMPS calculations in the thermodynamic limit for various interaction strengths $J$ (symbols).
For strong $J$, the peak position matches with the pitch angle of the classical-spin configuration obtained by perturbation theory (PT), i.e., from Eq.\ (\ref{eq:ptres2}), see the solid line.
}
\label{fig:thofphi}
\end{figure}

In the extreme strong-coupling limit $J \to \infty$, the unique ground state of the system consists of local singlets between the localized spin and the local conduction-electron spin at each site of the lattice. 
All excitations are gapped by an energy of the order of $J$ and consequently all two-point correlations exhibit an exponential decay with increasing distance. 
For the unfrustrated chain with $t_{2}=0$ ($\varphi = 0$), the short-range magnetic order is known \cite{TSU97b,SU99} to be antiferromagnetic.
This reflects itself in a well-defined peak of the spin-structure factor
\begin{equation}
S(Q) = \frac{1}{L} \sum_{ij}^{L} e^{iQ(R_{i} - R_{j})} \langle \ff S_{i} \ff S_{j} \rangle
\label{eq:ssf}
\end{equation}
with a peak maximum at the wave vector $Q=\pi$ and a width determined by the spin correlation length.

At strong $J$, the spin-structure factor is dominated by a single well-defined peak in the entire range $0 \leq \varphi \leq \frac{\pi}{2}$.
The evolution of the position of the peak maximum with increasing frustration $t_{2}$, i.e., $Q(\varphi)$, is displayed in Fig.\ \ref{fig:thofphi}. 
For small $\varphi$, the maximum remains at $Q=\pi$ until a critical $\varphi_{\rm c}(J)$ is reached. 
For $\varphi > \varphi_{\rm c}(J)$, the peak maximum smoothly shifts away from $Q=\pi$ and finally approaches $Q=\frac{\pi}{2}$ for $\varphi \to \frac{\pi}{2}$.
Hence, there is a sharp crossover from antiferromagnetic to incommensurate short-range magnetic order. 

The crossover point $\varphi_{\rm c}(J)$ is very weakly $J$-dependent in the strong-$J$ regime and almost perfectly matches the prediction $\varphi_{\rm c}(J\to \infty) = \varphi_{\infty} = \arctan(\frac{1}{2}) \approx 0.148 \pi$ of strong-coupling perturbation theory for the classical-spin Kondo lattice.
Furthermore, there is almost perfect agreement of the peak position $Q(\varphi)$ with the pitch angle $\theta(\varphi)$, as given by Eq.\ (\ref{eq:ptres2}), in the entire $\varphi$ range (see black line in Fig.\ \ref{fig:thofphi}).
Slight deviations that are visible on the scale used in the figure show up for $J=4t$ only and close to the ``critical'' point.
We conclude that the short-range spin correlations are purely classical. 
The quantum character of the spins does manifest itself in the long-distance limit though. 
Opposed to the classical-spin model, quantum fluctuations destroy the long-range AF or IC order. 
The quantum state is rather characterized by a finite spin gap $\Delta E_{\rm S}$ (see below) and, correspondingly, by exponentially decaying spin correlations on a scale given by a finite correlation length.

With decreasing $J$ the crossover point $\varphi_{\rm c}$ increases. 
The line $\varphi_{\rm c}(J)$ clearly separates two states with different (antiferromagnetic and incommensurate) short-range magnetic order down to $J \approx 0.9$ and comes close to touch the critical line for the dimerization transition; see the green line separating AF-SRO and IC-SRO in Fig.\ \ref{fig:dmrg-phase}.
This phase-diagram topology is almost the same as in the classical-spin case (cf.\ Fig.\ \ref{fig:pdclass} for comparison).
Essentially, there are two differences: 
First, in the classical-spin case, the triple point $(J_{\rm tri}, \varphi_{\rm tri}) \approx (4.1t, 0.188 \pi)$ is found at a much stronger interaction.
This is interpreted as being due to a destabilization of the spin dimerization caused by quantum fluctuations.
Second, in the quantum-spin case, and for $J$ very close to the dimerization transition, there is another peak developing in the spin-structure factor close to but larger than $\frac{\pi}{2}$. 
The emergence of this peak is accompanied by a disappearance of the crossover at $\varphi_{\rm c}(J)$ for $J \lesssim 0.9t$.
Namely, the peak structure at $Q=\pi$ first remains but becomes weak as compared to the new peak at $Q \gtrsim \frac{\pi}{2}$ and finally vanishes. 

The peak in $S(Q)$ at $Q \approx \frac{\pi}{2}$ (but $Q > \frac{\pi}{2}$) is interpreted as a precursor of the magnetic $Q=\frac{\pi}{2}$ spiral phase found at weaker $J$ (see the red transition line in the phase diagram Fig.\ \ref{fig:dmrg-phase}).
It is observed whenever $(J,\varphi)$ is close to the magnetic transition line.
There is, however, no {\em direct} transition from the gapped phase with magnetic short-range order to the gapless quasi-long-range ordered spiral phase. 
The magnetic phase transition is rather pre-empted by spin dimerization.

\subsection{Translational symmetry breaking}

\begin{figure*}
\includegraphics[width=1.75\columnwidth]{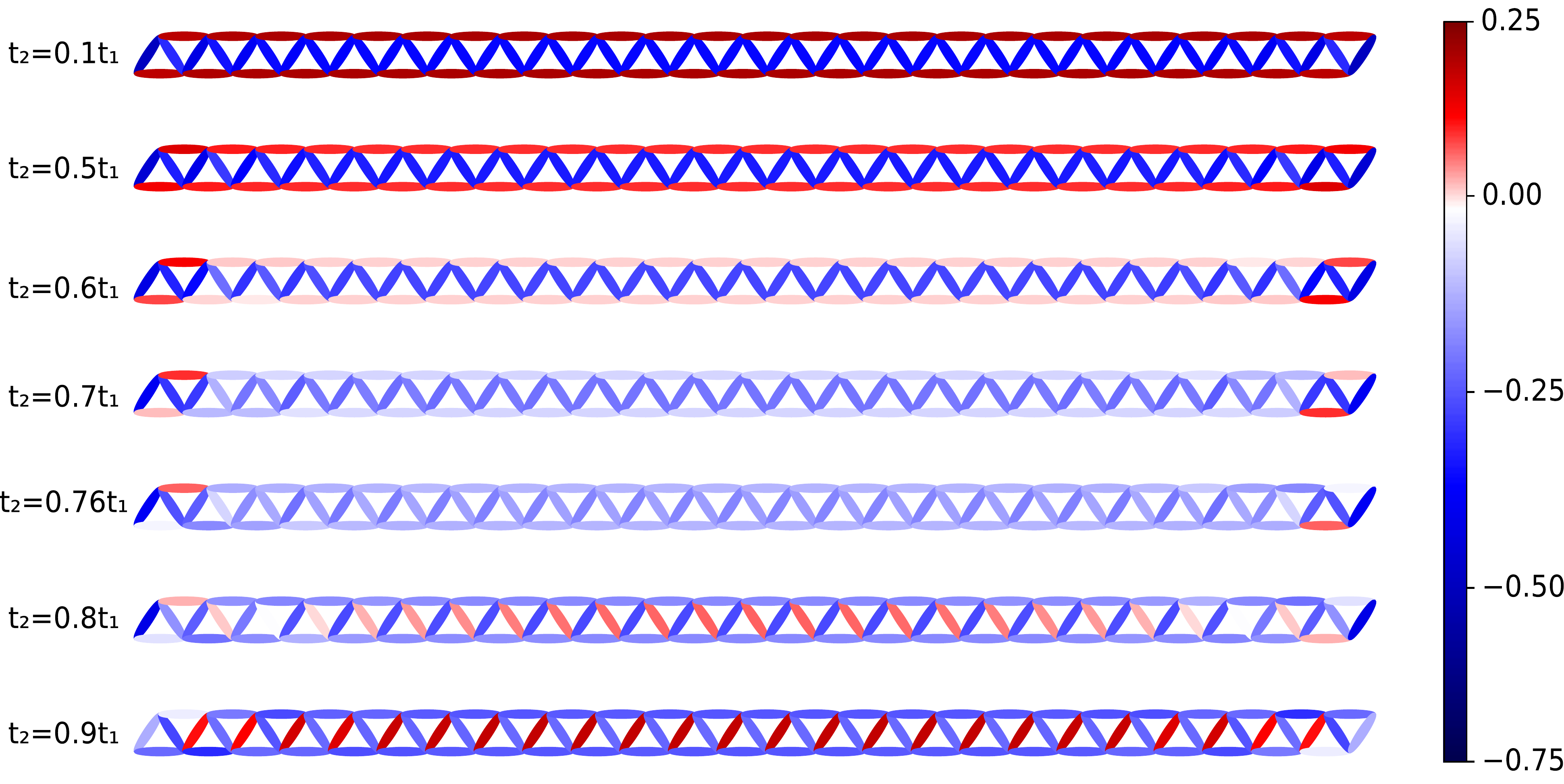}
\caption{
Short-range spin correlations $\langle \ff S_{i} \ff S_{j} \rangle$ (color code) as obtained from DMRG for a chain with $L=52$ sites at $J=0.7t_1$ and for various values of the next-nearest neighbor hopping $t_2$ as indicated.
}
\label{fig:short-range}
\end{figure*}

Dimerization is the simplest form of a spontaneous breaking of the discrete translational symmetries of a one-dimensional lattice. 
In a dimerized state, the nearest-neighbor correlation $\langle A_{i} A_{i+1} \rangle$ of a local observable $A_{i}$ depends on the site index $i$ and  alternates around the spatial average, i.e., the dimerization order parameter
\begin{equation}
   \langle A_{i-1} A_{i} \rangle - \langle A_{i} A_{i+1} \rangle = \mbox{const.} \times (-1)^{i}
\end{equation}
oscillates, while $\langle A_{i} \rangle = \langle A \rangle = \mbox{const}$, since each site $i$ belongs to both inequivalent bonds $(i-1,i)$ and $(i,i+1)$.
Here, we discuss spin dimerization with order parameter
\begin{equation}
   O_{D} = | \langle \ff S_{i} \ff S_{i+1} \rangle - \langle \ff S_{i+1} \ff S_{i+2} \rangle | \: . 
\label{eq:dim}
\end{equation}

The ground state of the unfrustrated ($t_2=0$) Kondo lattice at half-filling is homogeneous in the entire $J$-range with antiferromagnetic nearest-neighbor spin correlations $\langle \ff S_{i} \ff S_{i+1} \rangle$ while next-nearest neighbor correlations $\langle \ff S_{i} \ff S_{i+2} \rangle$ are ferromagnetic.
Contrary, as has been shown in Ref.\ \onlinecite{PRP18}, the Kondo model on the zigzag lattice with $t_2=t_1$ has weakly antiferromagnetic spin correlations on the rungs as well as for nearest neighbors along the legs of the ladder for strong $J$ and undergoes a spin-dimerization transition at $J = \jd \approx 0.89 t_1$.
For $J < \jd$, the spin correlations on the rungs alternate between different antiferromagnetic or even between anti- and ferromagnetic values.
Consequently, we expect a spin-dimerization transition for fixed $J$ when passing from $t_{2}=0$ to $t_{2}=t_{1}$.

Figure \ref{fig:short-range} demonstrates that this is indeed the case. 
It displays the nearest-neighbor spin correlations on the rungs and along the legs for at $J=0.7t_1$ and for various values of $t_2$. For weak frustration, $t_2=0.1t_1$ (top panel), we find strong antiferromagnetic correlations on the rungs and ferromagnetic correlations on the legs as for the unfrustrated limit $t_{2}=0$.
With increasing $t_2$ up to $t_{2}=0.7$, frustration of antiferromagnetic order gets more and more important and leads to weaker but still antiferromagnetic rung correlations. 
At the same time, the increasing hopping along the legs leads to a decrease of the ferromagnetic leg spin correlations and finally, for $t_{2}=0.7t_{1}$, to antiferromagnetic values $\langle \ff S_{i} \ff S_{i+2} \rangle < 0$.
The figure also demonstrates that edge effects, though clearly visible, do not affect the central region of the ladder for $L=52$.

At $t_2=0.76t_1$, the system has undergone the spin-dimerization transition.
There is a clearly alternating pattern between weaker and stronger antiferromagnetic correlations on the rungs which breaks the translational symmetry of the system. 
This dimerization pattern is even more apparent at $t_2=0.8t_1$, as rung correlations are now even oscillating between antiferro- and ferromagnetic values.
In the thermodynamic limit $L\to \infty$, the ground state would be twofold degenerate. 
For the finite system studied here ($L=52$) and for the particular geometry considered, the presence of the edges explicitly breaks the symmetry with respect to a reflection at the chain center and thus favors one of the two states.
At $t_{2}=0.9t_{1}$, the spin correlations form a structure reminiscent of unfrustrated antiferromagnetic correlations that would emerge when switching off the nearest-neighbor hopping on every second rung, thereby forming a simple bipartite lattice again.
In this way, the spin-dimerization alleviates magnetic frustration, and thus paves the way for quasi-long-range magnetic order
(see Sec.\ \ref{sec:qlro}).
Note that this is somewhat different but very much in the same spirit as the mechanism of partial Kondo screening (PKS) suggested for the frustrated Kondo lattice on the two-dimensional triangular lattice. \cite{MNYU10}

\begin{figure}[t]
\includegraphics[width=0.95\columnwidth]{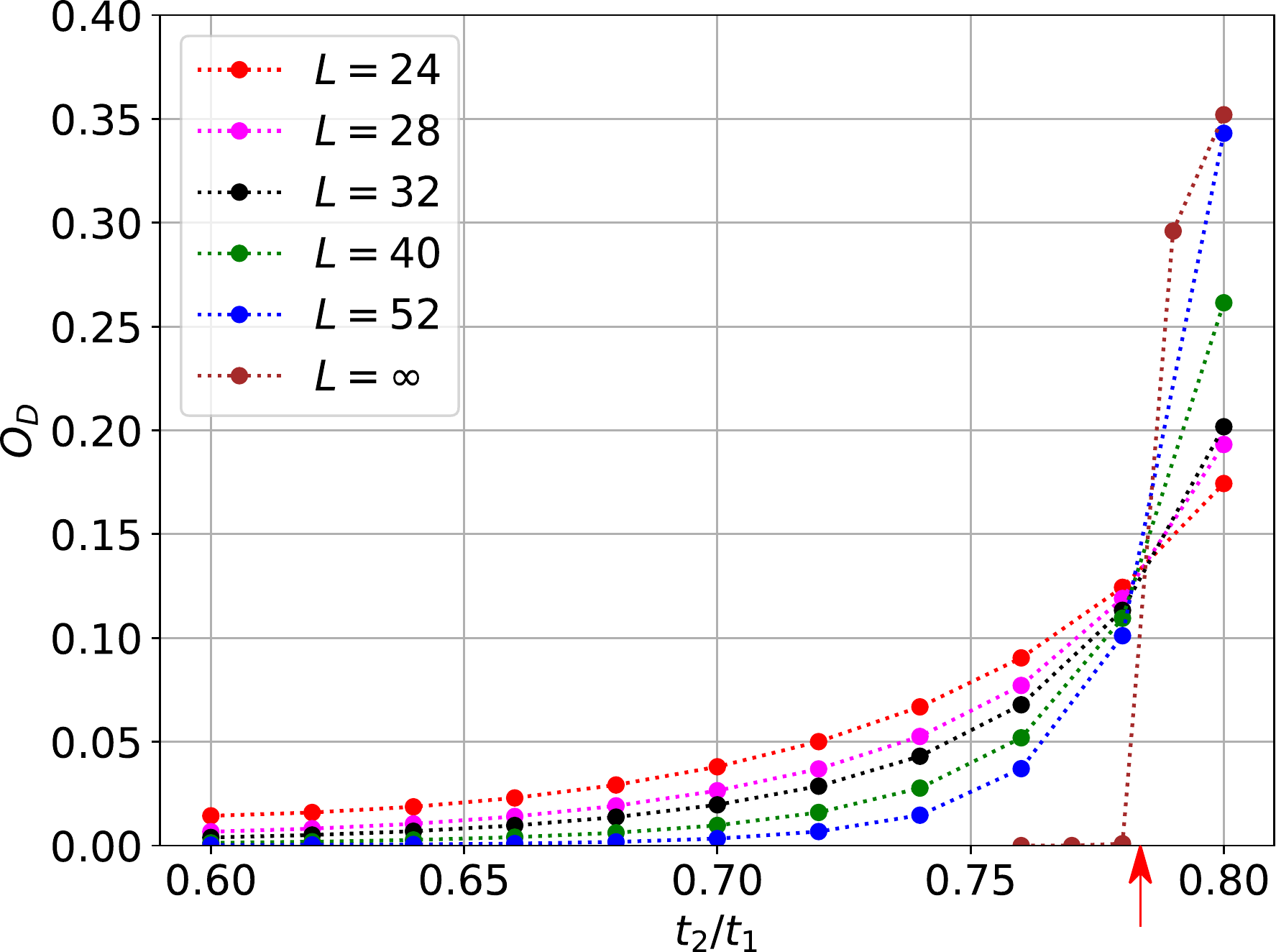}
\caption{
Order parameter for spin dimerization $O_{\rm D}$, Eq.\ (\ref{eq:dim}), as a function of $t_2/t_1$ for different $L$ at
$J=0.7t_1$, as obtained by DMRG. 
The $L=\infty$ values are obtained by the VUMPS algorithm after scaling to the infinite bond-dimension limit.
The red arrow marks the transition: $(t_{2} / t_{1})^{\rm (dim)}_{\rm c}\approx 0.785 t_{1} \pm 0.005 t_1$. 
This corresponds to $\varphi_{\rm c}^{\rm (dim)} \approx 0.67 \approx 0.21 \pi$ at $J \approx 0.55t$, see the phase diagram in Fig.\ \ref{fig:dmrg-phase}.
}
\label{fig:tcd}
\end{figure}

A precise location of the transition point can be achieved by studying the dimerization order parameter $O_{\rm D}$, computed via Eq.\ (\ref{eq:dim}) from the site-dependent nearest-neighbor correlations $\langle \ff S_{i} \ff S_{i+1} \rangle$. 
DMRG results for $O_{\rm D}$ are presented in Fig.\ \ref{fig:tcd} for several system sizes $L$. 
The transition point is expected to be located at the crossing of the curves obtained for different finite $L$. 
This yields a value $(t_{2} / t_{1})^{\rm (dim)}_{\rm c}\approx 0.79 t_1$, which is also consistent with additional VUMPS calculations working in the thermodynmic limit $L=\infty$. 
The latter, after scaling to the infinite bond-dimension limit, locate the transition point at $(t_{2} / t_{1})^{\rm (dim)}_{\rm c} \approx 0.785 t_{1} \pm 0.005 t_1$, corresponding to $\varphi_{\rm c}^{\rm (dim)} \approx 0.67 \approx 0.21 \pi$ at $J \approx 0.55t$, see the phase diagram Fig.\ \ref{fig:dmrg-phase}.
The VUMPS data shown in Fig.\ \ref{fig:tcd} suggest a first-order (or weakly first-order) transition, the order parameter $O_{\rm D}$ appears to jump at the transition point.
This is supported by the finite-size DMRG data when extrapolating to the $L\to \infty$ limit.

We have systematically computed the phase boundary for the transition to the dimerized state in the $J$-$\varphi$ plane.
There is a large region, where the ground state has a broken translational symmetry (see blue dots in Fig.\ \ref{fig:dmrg-phase}). 
We were able to trace the transition down to $J \approx 0.1t$. 
Generally, computations get much more involved in the weak-$J$ regime. 
Here, the physics is more and more dominated by RKKY-type long-range effective spin interactions which produce an increase of the entanglement entropy. 
For the weakly frustrated regime ($t_2 \ll t_1$), the situation is different. 
Here, calculations can be done for $J\sim 0.1t_1$ without difficulty, and the phase boundary of the spin-dimerized phase could be obtained accurately. 
{\em Within} the symmetry-broken phase (for $t_2\gtrsim0.5t_1$ at $J=0.1t_1$), however, the entanglement is considerably higher, so that computations become very challenging and prevent us from further systematic studies in this regime.

Starting from $\varphi = \pi/4$ (i.e., $t_{1}=t_{2}$) $\varphi_{c}(J)$ the phase boundary decreases with decreasing $J$. 
The data support a scenario with $\varphi_{\rm c}(J) \to \varphi_{0}  \approx 0.148 \pi$ for $J\to 0$, see the solid blue line.
This exactly recovers the result of the classical-spin model and coincides with the $\varphi$, at which the number of Fermi points in the noninteracting band structure changes.

Starting again from $\varphi = \pi/4$ but {\em increasing} $J$, the phase boundary $\varphi_{c}(J)$ first increases, reaches a maximum at $J\approx0.68t$, and  then bends back, see Fig.\ \ref{fig:dmrg-phase}.
This is important as the boundary {\em must} bend back, in fact. 
The dashed blue line indicates a possible further trend.
The reason is that it is impossible to have the transition line ending in a critical point $J=J_{\rm c}$ at $\varphi=\pi / 2$, i.e., in the decoupled-chain limit $t_{1}=0$.
Clearly, in this limit the ground state is nondegenerate and excitations are gapped for all $J>0$ so that nondegenerate perturbation theory with respect to the hopping term $t_{1}$ applies, excluding a dimerization transition along $t_{1}=0$.
Note that this is a qualitative difference to the phase diagram of the classical-spin model (see Fig.\ \ref{fig:pdclass}).
The important point is that in the classical-spin case the ground-state for $t_{1}=0$ is highly degenerate.
It is tempting to assume that the phase boundary finally ends at $\varphi (J=0) = \pi / 2$, since apart from $\varphi_{0}$ there is no further qualitative change in the noninteracting dispersion (see Fig.\ \ref{fig:fermi}).

\subsection{Quasi-long-range magnetic order}
\label{sec:qlro}

The unfrustrated model at $t_{2}=0$ exhibits short-range antiferromagnetic order with an exponential decay of the spin-correlation function for all $J>0$ and a finite energy gap to excited states.
For finite $t_{2}>0$ and with increasing $t_{2}$, antiferromagnetic correlations are more and more frustrated. 
At $t_{2}=t_{1}$, however, the ground state supports quasi-longe-range antiferromagnetic order at interactions weaker than the critical interaction $J^{\rm mag}_{\rm c}\approx0.84t_1$ (i.e., $J^{\rm mag}_{\rm c} \approx 0.594 t$, see Fig.\ \ref{fig:dmrg-phase}), as in known from our previous work. \cite{PRP18}
In view of the strong geometrical frustrations, this is quite surprising.
The magnetic state shows up on top of the spin dimerization, and the magnetic order is given by a $90^\circ$ spin spiral rather than a collinear antiferromagnetic state.
The quasi-long-range order is characterized by algebraically decaying spin correlations, by a vanishing spin gap, and by a spin-structure factor diverging at the wave vector $Q=\frac{\pi}{2}$ in the thermodynamic limit. 
Note that according to the Mermin-Wagner theorem, \cite{MW66} the presence of quantum fluctuations merely excludes {\em long-range} order for the ground state of one-dimensional systems.

Here, we like to check this result by studying the emergence of the spiral state for fixed $J$ with increasing $t_{2}$.
Furthermore, the magnetic phase boundary shall be tracted in the $J$-$\varphi$ plane to see whether the magnetic state is bound to the spin dimerization. 
To this end we compute the spin gap $\Delta E_{\rm S}$, which is defined as 
\begin{equation}
  \Delta E_{\rm S} = E(L,1) - E(L,0) 
\label{eq:sgap}
\end{equation}
where $E(N_{\rm tot},S_{\rm tot})$ is the ground-state energy in the sector of the Hilbert space with total particle number $N_{\rm tot}$ and total spin-quantum number $S_{\rm tot}$. 

\begin{figure}[t]
\includegraphics[width=0.95\linewidth]{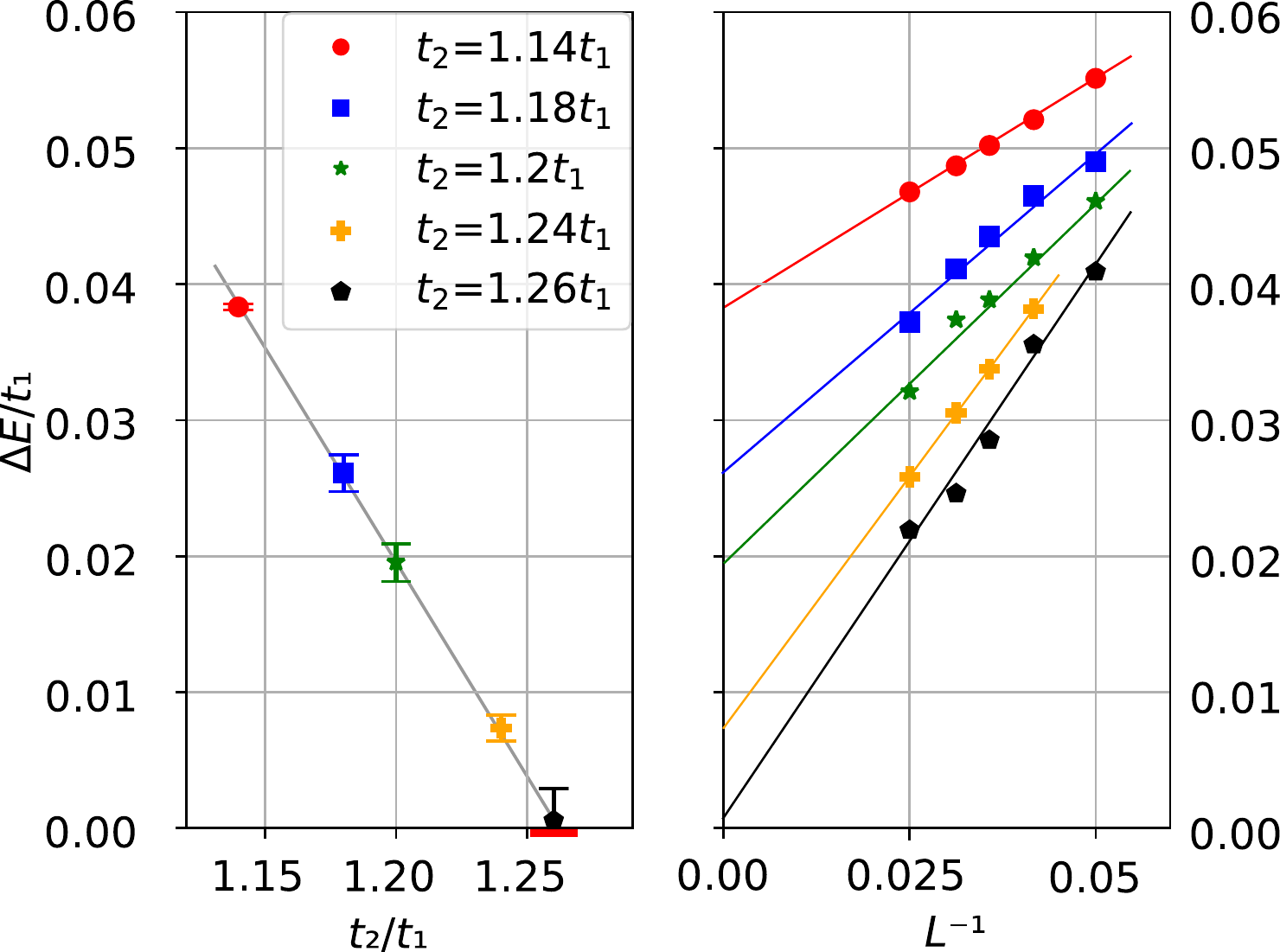}
\caption{
Spin gap $\Delta E_{\rm S}$ for $J=1.0 t_1$ and different $t_2/t_1$ as indicated. 
{\em Right:} $L$-dependence of the spin gap. 
Lines: linear fits $\Delta E_{\rm S}(L) - \Delta E_{\rm S}(\infty) \propto 1/L$. 
{\em Left:} $\Delta E_{\rm S}(\infty)$, obtained by extrapolation to the $1/L \to 0$ limit, as a function of $t_2/t_1$.
Bars: extrapolation error.
A linear fit of the $t_2$-dependence of $\Delta E_{\rm S}(\infty)$ (grey line) yields the critical value $t^{\rm mag}_{2,\rm c} = 1.26t_{1} \pm 0.03t_{1}$ (see red horizontal bar). 
This corresponds to $\varphi_{\rm c}^{\rm mag}(J) \approx 0.90 \approx 0.29 \pi$ at $J \approx 0.62t$, see the phase diagram in Fig.\ \ref{fig:dmrg-phase}. 
}
\label{fig:gap}
\end{figure}

Fig.\ \ref{fig:gap} displays the spin gap at $J=1.0t_1$ and as a function of $t_{2}$ in the critical regime. 
$\Delta E_{\rm S} = \Delta E_{\rm S}(L)$ is computed for several system sizes $L \le 40$ and extrapolated to the thermodynamical limit $L \to \infty$.
With increasing hopping along the legs of the ladder $t_{2}$, i.e., with increasing frustration, the spin gap decreases and finally closes at a critical value $t^{\rm mag}_{2,\rm c}$. 
In the vicinity of the transition, $\Delta E_{\rm S}(\infty)$ is an almost linear function of $t_{2}$, such that the critical hopping is obtained by linear extrapolation. We find $t^{\rm mag}_{2,\rm c} = 1.26t_{1} \pm 0.03t_{1}$ (see the red horizontal bar in Fig.\ \ref{fig:gap}), corresponding to $\varphi_{\rm c}^{\rm mag}(J) \approx 0.90 \approx 0.29 \pi$ at $J \approx 0.62t$ (see Fig.\ \ref{fig:dmrg-phase}).

\begin{figure}
\includegraphics[width=0.9\columnwidth]{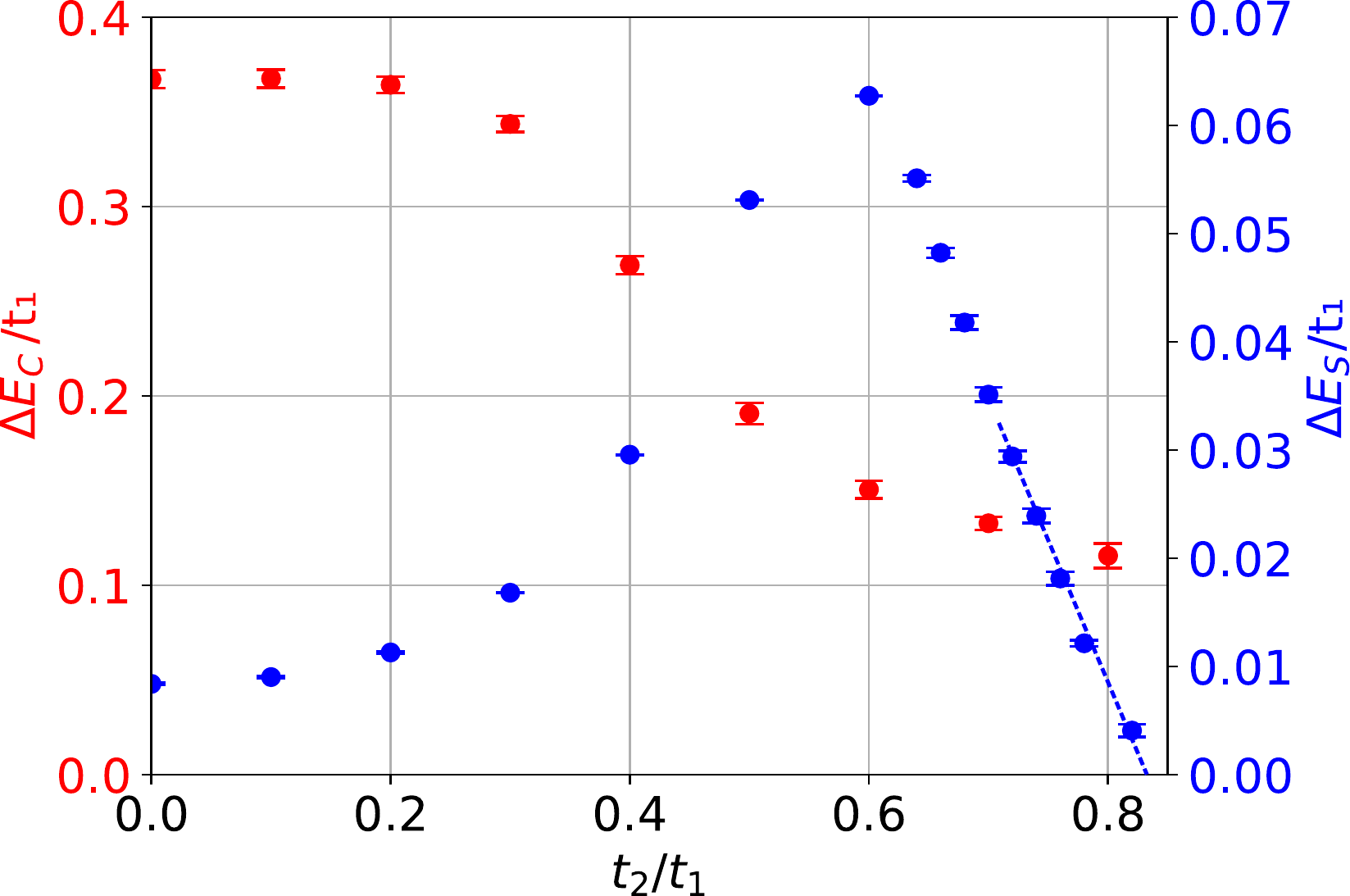}
\caption{
The spin gap $\Delta E_{\rm S}$ [see Eq.\ (\ref{eq:sgap})] and the charge gap $\Delta E_{\rm C}$ [see Eq.\ (\ref{eq:cgap})] as functions of next-nearest-neighbor hopping $t_2/t_1$ at $J=0.7t_1$. 
All values are obtained by extrapolation to the thermodynamic limit.
Bars: errors of the fit. 
Blue dotted line: linear fit $\Delta E_{\rm S} \propto \left( t_2 - t^{\rm mag}_{2,\rm c} \right)$.
The spin gap closes at $t^{\rm mag}_{2,\rm c} \approx0.83 t_{1} \pm 0.03t_1$.
This corresponds to $\varphi_{\rm c}^{\rm mag}(J) \approx 0.69 \approx 0.22 \pi$ at $J \approx 0.54t$, see the phase diagram in Fig.\ \ref{fig:dmrg-phase}.
}
\label{fig:gapSC}
\end{figure}

Fig. \ref{fig:gapSC} displays our data for the spin gap, extrapolated to the $L=\infty$ limit, for $J=0.7t_1$ as a function of $t_2/t_1$.
Here, we give an overview over the full $t_{2}$ range from $t_{2}=0$ up to the phase transition. 
The charge gap, 
\begin{equation}
  \Delta E_{\rm C} = \frac{E(L+2,0) + E(L-2,0) - 2E(L,0)}{2} \; , 
\label{eq:cgap}  
\end{equation}
is shown in addition. 

For the unfrustrated Kondo lattice at $t_{2}=0$, both the spin and the charge gap are finite, and the charge gap is almost an order of magnitude larger than the spin gap.
This is consistent with earlier DMRG studies. \cite{SU99} 
With increasing $t_{2}$, the charge gap decreases and appears to saturate at a finite value for the largest $t_{2}$ available.
This is the expected trend as this increases the itineracy of the electrons at fixed $J$.

The dependence of the spin gap on $t_{2}$, however, is unconventional. 
With increasing $t_{2}$, antiferromagnetic nearest-neighbor correlations are more and more frustrated. 
This explains the initial increase of $\Delta E_{\rm S}$.
However, the spin gap reaches a maximum at $t_2\approx0.6t_1$ and starts to decrease when further increasing $t_{2}$. 
Eventually, the spin gap closes at the critical value $t^{\rm mag}_{2, \rm c} \approx0.83 \pm 0.03t_1$, corresponding to $\varphi_{\rm c}^{\rm mag}(J) \approx 0.69 \approx 0.22 \pi$ at $J \approx 0.54t$ (see Fig.\ \ref{fig:dmrg-phase}).
This signals the onset of quasi-long-range magnetic order.

The phase boundary to the magnetic state has been computed for several values in the $J$-$\varphi$ plane, see red data points in Fig.\ \ref{fig:dmrg-phase}.
On the phase boundary, the spin-structure factor, Eq.\ (\ref{eq:ssf}), diverges at $Q=\frac{\pi}{2}$ as $L \to \infty$. 
Our data are well described by a logarithmic divergence $S(Q=\pi / 2) \sim \ln L$ corresponding to a decay of long-range spin correlations $\langle \ff S_{i+\Delta i} \ff S_{i} \rangle \propto e^{-iQ\Delta i}/\Delta i$, see Refs.\ \onlinecite{Lia90,LC91,SS93,KM96}. 
Hence, the order is characterized as a $90^{\circ}$ spin spiral.

As can be seen from Fig.\ \ref{fig:dmrg-phase}, the magnetic quasi-long-range order shows up {\em within} the spin-dimerized phase. 
The magnetic phase boundary is clearly distinct from but closely follows the boundary to the dimerized phase. 
This strongly corroborates our interpretation that the spin dimerization can be seen as a precursor of the magnetic state. 
The pattern of nearest-neighbor spin correlations in the dimerized phase, see Fig.\ \ref{fig:short-range} at $J=0.8t_{1}$ or $J=0.9t_{1}$, for example, comes close to a pattern formed on a bipartite lattice. 
Dimerization strongly alleviates the frustration and thus helps to build up longer-range spin correlations. 
As pointed out above, this is similar to the mechanism of partial Kondo screening (PKS). \cite{HUM11,MNYU10}

It is an open question how far the spiral magnetic order extends.
Either it exists down to the $J\to 0$ range or it forms an ``island'' in the $J$-$\varphi$ plane.
We performed calculations as shown in Fig.\ \ref{fig:gapSC} but for weaker $J$. 
Using DMRG, we were not able, however, to trace the magnetic phase boundary to the weak-$J$ regime. 
The spin gap gets progressively smaller with decreasing $J$ in the nonmagnetic state and is eventually dominated by an exponentially small energy scale, similar to the Kondo scale of the corresponding impurity problem. \cite{SU99} 
As a result, the DMRG computations get too costly to achieve sufficiently small errors.
Still we could verify that the spin gap is finite in the range from $t_{2}=0$ up to the onset of spin dimerization and down to $J=0.1 t_{1}$. 

VUMPS calculations in the weak-$J$ regime and {\em within} the dimerized phase do not give a conclusive answer either. 
Calculations suffer from a comparatively large standard deviation of the total energy per site, 
computed following Ref.\ \onlinecite{ZSVF+18}, which typically amounts to about 5\% of $t$ at bond dimension $m \approx 15,000$ (see Sec.\ \ref{sec:mps}).
This is not sufficient for reliable results. 
To give an example, for $t_{2}=t_{1}$ and at weak $J$, e.g., $J=0.1t$,  we find a state with extremely strong spin dimerization, which is close to a valence-bond solid and indications for which have been reported in Ref.\ \onlinecite{PRP18} already.
The state (i.e., one of two degenerate ground states) is characterized by almost perfect nearest-neighbor local-spin singlets $\langle \ff S_{i} \ff S_{i+1} \rangle \approx -0.75$ for, say, the even sites $i$ while $\langle \ff S_{i} \ff S_{i+1} \rangle \approx 0$ for odd $i$.
Beyond nearest neighbors, spin correlations $\langle \ff S_{i} \ff S_{j} \rangle$, are tiny (of the order of $10^{-3}$) and decay exponentially (at finite bond dimension $m$). 
The spin-structure factor is peaked at the wave vector $Q=\frac{\pi}{2}$ due to short-range spiral magnetic order on a length scale of several tens of lattice constants.
The interesting question is, whether there is quasi-long-range magnetic order.
However, extracting the correlation length by fitting the long-distance behavior of $S(Q)$ for several bond dimensions $m$ and extrapolation to $m=\infty$ is not possible since the results depend in an irregular way on $m$. 
Furthermore, in this parameter regime, there is an artificial dependence of the results on the choice of the size of the unit cell $k$. 
Using $k=2,4,6,8$, we find more complex valence-bond-solid-like states almost degenerate with, and at $k=4$ even stable as compared to the spin-dimerized one. 
VUMPS calculations using much higher bond dimensions ($m \gg 15,000$) would be necessary to get an acceptable variance, and conclusive results on the question of the quasi-long-range order and the stability of the valence-bond-like state at weak $J$.

\section{Discussion and conclusions}
\label{sec:dis}

Let us summarize and discuss where we are at in the Kondo-lattice problem: 
The limit $t_{2}=0$, i.e., the unfrustrated one-dimensional Kondo lattice at half-filling is well understood. 
The ground state is nondegenerate with gapped charge and spin excitations. \cite{TSU97b,SU99}
Spin correlations are antiferromagnetic and short ranged.
In the weak-$J$ limit, the spin gap is exponentially small. \cite{YW93,Tsv94,WLL93,JP81}
Nondegenerate perturbation theory smoothly connects the model with weak $t_{2}>0$ to the $t_{2}=0$ limit.
The same holds for the limit $t_{1}=0$, i.e., two decoupled Kondo chains. 
At arbitrary $J>0$ and weak $t_{1}>0$, the ground state is smoothly connected to the $t_{1}=0$ limit.
There is no quantum-phase transition in both limits.

For finite $t_{2}$ (and $t_{1}$), the system is magnetically frustrated. 
At a fixed ratio of the hopping $t_{2} / t_{1} = \tan \varphi$, frustration becomes more and more effective with decreasing $J$.
At fixed $J$, on the other hand, antiferromagnetic correlations along the legs tend to be less frustrated as compared to the correlations on the rungs. 
The purely geometrical reason is that there is only a single path involving two nearest-neighbor hops, which connects nearest neighbors along the legs, while there are two such paths, which connect nearest neighbors on the rungs.
Hence, frustration is expected to be strongest for $t_{2} < t_{1}$.
Note that the phase diagram is indeed asymmetric with respect to $\varphi = \frac{\pi}{4}$.

The asymmetry already shows up for strong $J$.
In the $J=\infty$ limit, the ground state is a trivial product of local Kondo singlets. 
At strong but finite $J$, however, we find a precisely defined boundary between a state with antiferromagnetic (AF) short-range spin correlations, for $0 \le \varphi \le \varphi_{\rm c}(J)$, and a state with incommensurate (IC) short-range order (for $\varphi_{\rm c}(J) \le \varphi \le \frac{\pi}{2}$). 
This is easily understood with the help of the classical-spin model. 
Due to the absence of quantum fluctuations, the ground state exhibits {\em long-range} AF or IC magnetic order, and the phase transition between the AF and the IC phase perfectly matches with the shift of the maximum of the spin-structure factor from $Q=\pi$ for $\varphi < \varphi_{\rm c}(J)$ to $Q<\pi$ for $\varphi > \varphi_{\rm c}(J)$ in the quantum-spin case.
In this regard, spin correlations in the strong-$J$ regime are almost classical. 
For $J \to \infty$, the $\varphi$-dependence of the pitch angle of the incommensurate magnetic order can be determined analytically by mapping onto a classical Heisenberg model on the zigzag ladder.
At $\varphi = \pi /4$ ($t_{1}=t_{2}$), we recover the pitch angle $\theta \approx 104.5^{\circ}$ on the zigzag ladder, \cite{Dub16} to be compared with the $120^{\circ}$ configuration in the classical Heisenberg model on the triangular lattice, for example.

The classical-spin model also explains the occurrence of spin dimerization. 
The dimerized phase, characterized by the pitch angle $\theta = \pi/2$ and an alternating angle $\theta \pm \Delta \theta$ with $\Delta \theta=\pi /2$ between classical spins on the rungs, has minimal ground-state energy in the weak-$J$-weak-$t_{1}$ range of the phase diagram, limited by a boundary $\varphi_{\rm c}^{\rm (dim)}(J)$.
Dimerization is a nonperturbative phenomenon. 
For $J \to 0$, we find that standard RKKY theory does not recover spin dimerization, at least not at order $J^{2}$, even though the phase boundary ends at $J=0$ for a finite hopping ratio: $\varphi_{\rm c}^{\rm (dim)}(J) \to \varphi_{0} = \arctan(\frac 12)$ for $J\to 0$.
Nevertheless, it is obvious that in the $J\to 0$ limit the spin dimerization is caused by the Lifschitz transition at $\varphi_{0}$, where  the number of Fermi points in the noninteracting band structure changes.
Is is noteworthy that $\varphi_{0}$ coincides with $\varphi_{\infty} = \varphi_{\rm c}^{\rm (dim)}(J=\infty)$. 
However, this must be seen as a coincidental match; for example, the wave vectors of the incommensurate phase for strong $J$ do not correspond to nesting vectors connecting noninteracting Fermi points.

The phase diagram for the classical-spin model is characterized by the boundary $\varphi_{\rm c}^{\rm (dim)}(J)$, up to which the dimerized phase is existing, starting from the weak-$J$-weak-$t_{1}$ limit, and by the line $\varphi_{\rm c}(J)$, separating AF and IC long-range magnetic order for strong $J$. 
The latter hits the dimerization boundary at a ``triple'' point, given by $J_{\rm tri} \approx 4.1t$ and $\varphi_{\rm tri} = \varphi_{c}(J_{\rm tri}) = \varphi^{\rm (dim)}_{c}(J_{\rm tri}) \approx 0.188 \pi$. 
Transitions between the AF and the IC phase are continuous, while transitions between the AF or IC phase and the dimerized phase are discontinuous.

This phase diagram is constitutive for the quantum-spin case but there are important additional effects of quantum fluctuations. 
First, quantum fluctuations destroy the long-range AF and IC magnetic order leaving, however, a clear separation between AF and IC  short-range ordered states.
Second, spin dimerization is strongly suppressed by quantum fluctuations, such that the spin-dimerized phase extends towards much weaker, an order of magnitude smaller interaction strengths $J$ only.
Third, the line $\varphi_{\rm c}^{\rm (dim)}(J)$ cannot terminate at a finite critical $J$ on the $t_{1}=0$ axis, opposed to the classical-spin case, since the ground state is unique and fully gapped, unlike the classical-spin case where the global SO(3) spin-rotation symmetry leads to an infinite ground-state degeneracy.
Apart from $\varphi_{0}$, which marks the termination point of $\varphi_{\rm c}^{\rm (dim)}(J)$ for $J \to 0$, not only in the classical but also in the quantum-spin case, there is no second critical point of the noninteracting bandstructure in the range $0 < \varphi < \frac{\pi}{2}$, so that it seems likely that $\varphi_{\rm c}^{\rm (dim)}(J) \to \frac{\pi}{2}$ for $J\to 0$ represents the second termination point on the $J=0$ axis.
Fourth, there is no well-defined triple point in the quantum-spin case.
With decreasing $J$, the boundary $\varphi_{\rm c}^{\rm (dim)}(J)$ becomes less and less well defined, and a second peak grows in the spin-structure factor at $Q \gtrsim \frac{\pi}{2}$, which is a precursor of the spiral ($Q=\frac{\pi}{2}$) phase at still weaker $J$.

An important difference between the Kondo model on the zigzag ladder for classical and for quantum spins is the different kind of how the system deals with the magnetic frustration. 
For strong $J$ and sufficiently strong $t_{2}$, i.e., in the moderately frustrated regime, incommensurate (long- or short-range) order represents the preferred compromise in both cases.
While the ground state of the classical-spin model supports long-range magnetic order in the entire $J$-$\varphi$ plane with different collinear or noncollinear magnetic structures, quantum fluctuations entirely destroy magnetic long-range order in the quantum-spin case. 
It is surprising, however, that quasi-long-range magnetic order with algebraic decay of the spin correlations is realized in a parameter regime (weak $J$ and strong $t_{2}$) of the quantum-spin model where frustration is expected to be dominant. 
Frustration in a way appears to {\em favor} magnetic order as far as possible to be consistent with the Mermin-Wagner theorem. 
To understand this effect, it is instructive to see that the quasi-long-range spiral ($Q=\frac{\pi}{2}$) order, SP-QLRO, only shows up {\em within} the spin-dimerized phase and that the phase boundary of the SP-QLRO phase closely follows the boundary for the onset of dimerization.
Spin dimerization is thus the main reaction of the system to magnetic frustration in this parameter regime while magnetic order is favored as a secondary effect.
Hence, SP-QLRO appears as an epiphenomenon that necessarily requires the alleviation of frustration due to a dimerized spin structure which mimics an unfrustrated bipartite lattice.

At weak $J$, one usually expects a spin-only Heisenberg model as the appropriate low-energy theory, at least if there is magnetic ordering. 
RKKY theory, \cite{RK54,Kas56,Yos57} however, predicts an oscillatory effective spin coupling which is long-ranged, $J_{ij} \propto 1/|i-j|$, such that a numerical solution is not straightforward.
Our DMRG and VUMPS calculations for the full Kondo lattice in the weak-$J$ limit and for weak next-nearest-neighbor hopping $t_{2}$, such that the system is still in its homogeneous phase with full translational symmetry ($\varphi < \varphi_{\rm c}^{\rm (dim)}(J)$), have shown that the system, in this parameter regime, has a unique ground state and at the same time a finite spin (and finite charge) gap.
Arguing with the Lieb-Schultz-Mattis (LSM) theorem, \cite{LSM61} this implies that the physics cannot be not captured by an effective spin-$1/2$ model, since the excitation spectrum would have to be gapless in this case.

There is also direct evidence that RKKY theory is not appropriate in our case.
For the classical-spin variant of the Kondo model, we have seen that the RKKY coupling constants $J_{\rm RKKY}(k)$ diverge at the wave vector $k=\pi$ or at $k=\frac{\pi}{2}$, depending on $\varphi$.
Furthermore, the RKKY theory does not capture the spin-dimerization transition at finite $L$ in the limit $J \to 0$, and in the thermodynamic limit $L \to \infty$, the convergence radius even shrinks to $J=0$.
On the other hand, for the quantum-spin Kondo lattice with $t_{2}=0$, previous studies \cite{YW93,Tsv94,WLL93,JP81,TSU97b,SU99} have shown that the spin gap is finite but exponentially small in $J$. 
Again this suggests that the low-energy physics is not RKKY-like but rather reminiscent of the Kondo effect.
Our calculations for weak $J$ and $\varphi < \varphi_{\rm c}^{\rm (dim)}(J)$ support this picture.

For stronger frustration in the weak-coupling regime, i.e., for $\varphi > \varphi_{\rm c}^{\rm (dim)}(J)$ and weak $J$, we find a spin-dimerized phase with a doubly degenerate ground state. 
Hence, given that an effective spin-$1/2$ model was applicable, the LSM theorem would no longer require a gapless spectrum. 
This scenario is well known from the antiferromagnetic $J_{1}$-$J_{2}$ spin-$1/2$ Heisenberg chain beyond the Majumdar-Ghosh point.\cite{MG69} 
For sufficiently strong $J_{2}$, this model exhibits spin dimerization as well. 
Furthermore, the ground state is two-fold degenerate, and spin excitations are gapped.
It is important to note, however, that the transition point at $\varphi_{\rm c}^{\rm (dim)}(J)$ in the zigzag Kondo lattice is not at all comparable to the Majumdar-Gosh point, since in the Kondo-lattice case spin excitations are gapped on both sides of the transition.

It is an open question whether the magnetic quasi-long-range ordered phase with gapless spin excitations, found at intermediate coupling strengths and on top of the spin dimerization, persists down to the $J\to 0$ limit.
Again, a two-fold degenerate ground state together with a gapless spectrum of spin excitations is not in the spirit of the LSM theorem, such that it is tempting to argue that the magnetic phase must form an ``island'' in the $J$-$\varphi$ plane and should not extend down to the $J=0$ axis. 
Also this argument would require, however, that the effective low-energy physics was correctly described by an effective spin-$1/2$ model, as suggested by RKKY theory.

Another conceivable idea is that the effective spin-only theory is different from RKKY in this parameter range and essentially involves spins with $S=1$, and possibly an admixture of $S=0$ singlets, since the unit cell of the spin-dimerized state consists of two sites.
In this case, a magnetic phase transition within the spin-dimerized state for $J \to 0$ at $\varphi_{c}$, closing the spin gap, appears unconventional, too, as it falls into the class of integer-spin Heisenberg chains. \cite{Hal83a,Hal83b}

We conclude that the weak-$J$ limit of the frustrated Kondo lattice in one dimension is not yet fully understood.
Apart from computationally demanding more systematic DMRG studies of the zigzag Kondo lattice in the weak-$J$ regime, it would be highly interesting to tackle the RKKY problem directly, i.e., to compute the RKKY couplings and to solve the resulting quantum-spin model with long-range spin interactions numerically.

\acknowledgments
This work has been supported by the Deutsche Forschungsgemeinschaft through the Sonderforschungsbereich 925 (project B5).

\appendix

\section{Strong-coupling perturbation theory for classical spins}
\label{sec:pt}

For $J\gg t$, we write $H = H_0 + H_1$ with
\ba
  H_0 = J \sum_{i} \ff s_{i} \ff S_{i} \: , \; 
  H_1 = \sum_{ij\sigma} t_{ij} c^\dagger_{i\sigma} c_{j\sigma} \: 
\ea
and apply nondegenerate perturbation theory for a given configuration of the classical spins with length $S=1/2$, as introduced in Sec.\ \ref{sec:pertstrong}. 
The unperturbed energy of the ground state $|0\rangle$ is $E_0 =  -LJ/4$. 
Assuming $t_{ii}=0$, the first-order correction vanishes.

At second order, the excited states $| n \rangle$ contributing to the $n$-sum in Eq.\ (\ref{eq:eofspt}) are given by tensor products of $L-2$ atomic states with a singly occupied electronic orbital and two atomic states with empty and doubly occupied orbitals. 
Hence, the excitation energy is $E_n - E_0 = - J(L-2)/4   + JL/4  = J/2$.

To compute the matrix element $\matrixel{0}{H_1}{n}$ or $\matrixel{0}{c^\dagger_{i\sigma} c_{j\sigma}}{n}$ for a given classical-spin configuration, we use coherent states 
\ba
|\theta,\phi\rangle = \begin{pmatrix}e^{i\phi}\cos(\theta/2)\\\sin(\theta/2)\end{pmatrix} \: , 
\ea
with $\phi=0$ for all sites $i$, as the spins are assumed to lie in the $x$-$z$-plane. 
A classical spin configuration is described by the set of polar angles $(\theta_{1}, ..., \theta_{L})$, and the unperturbed ground state can be written as $\ket{0} = \ket{\theta_1} \otimes ... \otimes \ket{\theta_L}$
For a given pair of sites $i,j$, we have:
\ba
\langle 0 | \, c^\dagger_{i\sigma} c_{j\sigma} \,| n \rangle 
= 
\langle \theta_i |\langle \theta_j |\, c^\dagger_{i\sigma}c_{j\sigma} \, | v \rangle_i c^\dagger_{j\sigma}
c^\dagger_{j-\sigma} | v \rangle_j
\ea
where $\ket{v}_{i}$ is the vacuum state at site $i$.
Hence, 
\ba
\sum_\sigma \langle 0 | \, c^\dagger_{i\sigma}c_{j\sigma} \,| n \rangle 
&=& \langle \theta_i |\uparrow\rangle_i \langle \theta_j |\downarrow\rangle_j
- \langle \theta_i |\downarrow\rangle_i \langle \theta_j |\uparrow\rangle_j
\nonumber \\ &=& \cos(\dfrac{\theta_i}{2})\sin(\frac{\theta_j}{2}) - ( i \leftrightarrow j)
\nonumber \\
&=& \sin(\frac{\theta_{ij}}{2})
\ea
where $\theta_{ij}$ denotes the angle between spins at sites $i$ and $j$. 
For nearest neighbors on a rung $\theta_{ij} = \theta \pm \Delta\theta$ while for nearest neighbors along a leg $\theta_{ij} = 2 \theta$, see Fig.\ \ref{fig:theta}.
This yields
\ba
&& \sum_{n \ne 0} \frac{|\langle 0 | H_{1} | n \rangle|^{2}}{E_{0}-E_{n}}
= 
-\frac{2}{J} \sum_{ij}^{i\ne j} t_{ij}^2 \, \sin^{2}(\theta_{ij}/2)
\nonumber \\ && =
- \frac{2 t_{1}^{2}}{J} L
\left( \sin^{2} \frac{\theta + \Delta\theta}{2}
+ \sin^{2}\frac{\theta - \Delta\theta}{2} \right)
- \frac{4t_{2}^{2}}{J} L \sin^{2}\theta
\nonumber \\ && =
- \frac{2 t_{1}^{2}}{J} L
- \frac{2 t_{2}^{2}}{J} L
+ \frac{2 t_{1}^{2}}{J} L
\cos \theta \, \cos \Delta \theta 
+ \frac{2t_{2}^{2}}{J} L \cos 2\theta \: .
\nonumber \\
\ea
As shown in Appendix \ref{sec:heis}, the energy for the classical $J_1$-$J_2$ Heisenberg model with $|\ff S_{i}|=1/2$ on the zigzag ladder is 
\be
E_{\rm Heis}(\theta,\Delta\theta) 
= \frac{L}{4} 
\left(  
J_1 \cos{\theta} \cos{\Delta\theta} 
+ 
J_2 \cos{2\theta} 
\right) \: . 
\ee
Hence, with 
\be
J_1 = \frac{8t_2^2}{J} \; , \quad
J_2 = \frac{8t_2^2}{J}
\label{eq:jeffs}
\ee
and with 
\ba
\Delta E = L \left(-\frac14 J - \frac{2t_1^2}{J} - \frac{2t_2^2}{J} \right)
\label{eq:offset}
\ea
we finally get
\be
E(\theta,\Delta\theta) 
=
\Delta E
+
E_{\rm Heis}(\theta,\Delta\theta) 
+
\ca O \left(\frac{t^3}{J^3}\right)
\: .
\label{eq:klmheis}
\ee

\section{Heisenberg model on the zigzag ladder}
\label{sec:heis}

Consider the Heisenberg model on the zigzag ladder with classical spins of length $|\ff S_{i}|=1/2$ and exchange interactions $J_{1}$ and $J_{2}$ between nearest neighbors along the rungs and along the legs, respectively. 
The energy of a spin configuration, parameterized via $\theta$ and $\Delta \theta$ is: 
\ba
E& =& 
\frac{1}{2}
\sum_{ij} J_{ij} \ff {S}_i \ff {S}_j 
\nonumber \\
&=& 
\frac{L}{8}
\left( J_{1} \cos(\theta + \Delta \theta) + J_{1} \cos(\theta - \Delta \theta) + 2J_{2} \cos (2\theta) \right)
\nonumber \\
&=&
\frac{L}{4}
\left( J_{1} \cos(\theta)  \cos(\Delta \theta) + J_{2} \cos (2\theta) \right) \: .
\ea
A minimum of the function $E=E_{\rm Heis.}(\theta,\Delta \theta)$ requires vanishing partial derivatives with respect to $\theta$ and $\Delta \theta$. 
This means:
\be
  \sin{\theta} \left( J_1 \cos{\Delta\theta} + 4J_2 \cos{\theta}  \right) = 0
\label{partialtheta}
\ee
and 
\be
  \cos{\theta} \sin{\Delta\theta} = 0 \: .
\label{eq:partialdeltatheta}
\ee

These conditions can be satisfied with (i) $\theta$ and $\Delta \theta$ being integer multiples of $\pi$.
There is a stable minimum if $J_{1} > 4 J_{2}$.
Another set of solutions (ii) is obtained by setting $\theta$ and $\Delta \theta$ to $\pi /2$ plus arbitrary integer multiples of $\pi$.
However, for arbitrary $J_{1}, J_{2}$ this is a saddle only.
The remaining case (iii) is 
\be
  \theta =  \arccos{\left(- \frac{J_1}{4 J_2}\right)}  \; , \quad \Delta\theta =  0
  \label{eq:sol3}
\ee
and yields a minimum if $J_{1} < 4 J_{2}$.

\section{Analysis of finite-size effects}
\label{sec:fs}

For the Kondo lattice we have $t_{2}/t_{1} = \tan\varphi$, such that for $t_{1} < 2 t_{2}$ and strong $J$ the optimal spin configuration is given by $\Delta \theta = 0$ and
\be
  \theta = \arccos \left( -\frac{1}{4\tan^{2}(\varphi)} \right) \: ,
\ee
see Eqs.\ (\ref{eq:jeffs}) and (\ref{eq:sol3}).

For the infinite system, $\theta$ must be considered as a continuous variable which smoothly approaches $\pi / 2$ from above as $\varphi \to \pi/2$ from below. 
For the actual numerical calculations, performed at finite $L$ and using periodic boundary conditions, $\theta$ runs on a discrete $\theta$-grid with spacings of $2\pi/L$.
As $\varphi \to \pi/2$ from below, the optimal $\theta$ jumps from $\theta = \pi/2 +2\pi / L$ to $\theta = \pi /2$ at a certain value $\varphi = \varphi_{L} < \pi / 2$. 
For large $L$, we can determine $\varphi_{L}$ from
\be
  \frac{\pi}{2} + \frac{\pi}{L} = \arccos \left( -\frac{1}{4\tan^{2}(\varphi_{L})} \right) \: .
\ee
Solving for $\varphi_{L}$ yields
\be
  \varphi_L = \arctan\left(\frac{1}{\sqrt{4\sin(\pi/L)}}\right) \: .
\label{eq:varcrit}
\ee
The convergence of $\varphi_{L}$ to $\varphi_{\infty}$ is rather slow.
System sizes of about $L \approx 50,000$ are necessary to determine $\varphi$ close to $\pi / 2$ with an accuracy of better than 1\%.


\begin{thebibliography}{39}
\expandafter\ifx\csname natexlab\endcsname\relax\def\natexlab#1{#1}\fi
\expandafter\ifx\csname bibnamefont\endcsname\relax
  \def\bibnamefont#1{#1}\fi
\expandafter\ifx\csname bibfnamefont\endcsname\relax
  \def\bibfnamefont#1{#1}\fi
\expandafter\ifx\csname citenamefont\endcsname\relax
  \def\citenamefont#1{#1}\fi
\expandafter\ifx\csname url\endcsname\relax
  \def\url#1{\texttt{#1}}\fi
\expandafter\ifx\csname urlprefix\endcsname\relax\def\urlprefix{URL }\fi
\providecommand{\bibinfo}[2]{#2}
\providecommand{\eprint}[2][]{\url{#2}}

\bibitem[{\citenamefont{Tsunetsugu et~al.}(1997)\citenamefont{Tsunetsugu,
  Sigrist, and Ueda}}]{TSU97b}
\bibinfo{author}{\bibfnamefont{H.}~\bibnamefont{Tsunetsugu}},
  \bibinfo{author}{\bibfnamefont{M.}~\bibnamefont{Sigrist}}, \bibnamefont{and}
  \bibinfo{author}{\bibfnamefont{K.}~\bibnamefont{Ueda}},
  \bibinfo{journal}{Rev. Mod. Phys.} \textbf{\bibinfo{volume}{69}},
  \bibinfo{pages}{809} (\bibinfo{year}{1997}).

\bibitem[{\citenamefont{Hewson}(1993)}]{Hew93}
\bibinfo{author}{\bibfnamefont{A.~C.} \bibnamefont{Hewson}},
  \emph{\bibinfo{title}{The Kondo Problem to Heavy Fermions}}
  (\bibinfo{publisher}{Cambridge University Press},
  \bibinfo{address}{Cambridge}, \bibinfo{year}{1993}).

\bibitem[{\citenamefont{Ruderman and Kittel}(1954)}]{RK54}
\bibinfo{author}{\bibfnamefont{M.~A.} \bibnamefont{Ruderman}} \bibnamefont{and}
  \bibinfo{author}{\bibfnamefont{C.}~\bibnamefont{Kittel}},
  \bibinfo{journal}{Phys. Rev.} \textbf{\bibinfo{volume}{96}},
  \bibinfo{pages}{99} (\bibinfo{year}{1954}).

\bibitem[{\citenamefont{Kasuya}(1956)}]{Kas56}
\bibinfo{author}{\bibfnamefont{T.}~\bibnamefont{Kasuya}},
  \bibinfo{journal}{Prog. Theor. Phys.} \textbf{\bibinfo{volume}{16}},
  \bibinfo{pages}{45} (\bibinfo{year}{1956}).

\bibitem[{\citenamefont{Yosida}(1957)}]{Yos57}
\bibinfo{author}{\bibfnamefont{K.}~\bibnamefont{Yosida}},
  \bibinfo{journal}{Phys. Rev.} \textbf{\bibinfo{volume}{106}},
  \bibinfo{pages}{893} (\bibinfo{year}{1957}).

\bibitem[{\citenamefont{Doniach}(1977)}]{Don77}
\bibinfo{author}{\bibfnamefont{S.}~\bibnamefont{Doniach}},
  \bibinfo{journal}{Physica B+C} \textbf{\bibinfo{volume}{91}},
  \bibinfo{pages}{231} (\bibinfo{year}{1977}).

\bibitem[{\citenamefont{Shibata and Ueda}(1999)}]{SU99}
\bibinfo{author}{\bibfnamefont{N.}~\bibnamefont{Shibata}} \bibnamefont{and}
  \bibinfo{author}{\bibfnamefont{K.}~\bibnamefont{Ueda}}, \bibinfo{journal}{J.
  Phys.: Condens. Matter} \textbf{\bibinfo{volume}{11}}, \bibinfo{pages}{R1}
  (\bibinfo{year}{1999}).

\bibitem[{\citenamefont{Shen}(1996)}]{She96}
\bibinfo{author}{\bibfnamefont{S.-Q.} \bibnamefont{Shen}},
  \bibinfo{journal}{Phys. Rev. B} \textbf{\bibinfo{volume}{53}},
  \bibinfo{pages}{14252} (\bibinfo{year}{1996}).

\bibitem[{\citenamefont{Peschke et~al.}(2018)\citenamefont{Peschke, Rausch, and
  Potthoff}}]{PRP18}
\bibinfo{author}{\bibfnamefont{M.}~\bibnamefont{Peschke}},
  \bibinfo{author}{\bibfnamefont{R.}~\bibnamefont{Rausch}}, \bibnamefont{and}
  \bibinfo{author}{\bibfnamefont{M.}~\bibnamefont{Potthoff}},
  \bibinfo{journal}{Phys. Rev. B} \textbf{\bibinfo{volume}{97}},
  \bibinfo{pages}{115124} (\bibinfo{year}{2018}).

\bibitem[{\citenamefont{Motome et~al.}(2010)\citenamefont{Motome, Nakamikawa,
  Yamaji, and Udagawa}}]{MNYU10}
\bibinfo{author}{\bibfnamefont{Y.}~\bibnamefont{Motome}},
  \bibinfo{author}{\bibfnamefont{K.}~\bibnamefont{Nakamikawa}},
  \bibinfo{author}{\bibfnamefont{Y.}~\bibnamefont{Yamaji}}, \bibnamefont{and}
  \bibinfo{author}{\bibfnamefont{M.}~\bibnamefont{Udagawa}},
  \bibinfo{journal}{Phys. Rev. Lett.} \textbf{\bibinfo{volume}{105}},
  \bibinfo{pages}{036403} (\bibinfo{year}{2010}).

\bibitem[{\citenamefont{Sato et~al.}(2018)\citenamefont{Sato, Assaad, and
  Grover}}]{SAG18}
\bibinfo{author}{\bibfnamefont{T.}~\bibnamefont{Sato}},
  \bibinfo{author}{\bibfnamefont{F.~F.} \bibnamefont{Assaad}},
  \bibnamefont{and} \bibinfo{author}{\bibfnamefont{T.}~\bibnamefont{Grover}},
  \bibinfo{journal}{Phys. Rev. Lett.} \textbf{\bibinfo{volume}{120}},
  \bibinfo{pages}{107201} (\bibinfo{year}{2018}).

\bibitem[{\citenamefont{Aulbach et~al.}(2015)\citenamefont{Aulbach, Assaad, and
  Potthoff}}]{AAP15}
\bibinfo{author}{\bibfnamefont{M.~W.} \bibnamefont{Aulbach}},
  \bibinfo{author}{\bibfnamefont{F.~F.} \bibnamefont{Assaad}},
  \bibnamefont{and} \bibinfo{author}{\bibfnamefont{M.}~\bibnamefont{Potthoff}},
  \bibinfo{journal}{Phys. Rev. B} \textbf{\bibinfo{volume}{92}},
  \bibinfo{pages}{235131} (\bibinfo{year}{2015}).

\bibitem[{\citenamefont{Hayami et~al.}(2011)\citenamefont{Hayami, Udagawa, and
  Motome}}]{HUM11}
\bibinfo{author}{\bibfnamefont{S.}~\bibnamefont{Hayami}},
  \bibinfo{author}{\bibfnamefont{M.}~\bibnamefont{Udagawa}}, \bibnamefont{and}
  \bibinfo{author}{\bibfnamefont{Y.}~\bibnamefont{Motome}},
  \bibinfo{journal}{J. Phys. Soc. Jpn.} \textbf{\bibinfo{volume}{80}},
  \bibinfo{pages}{073704} (\bibinfo{year}{2011}).

\bibitem[{\citenamefont{Rausch and Potthoff}(2016)}]{RP16}
\bibinfo{author}{\bibfnamefont{R.}~\bibnamefont{Rausch}} \bibnamefont{and}
  \bibinfo{author}{\bibfnamefont{M.}~\bibnamefont{Potthoff}},
  \bibinfo{journal}{New J. Phys.} \textbf{\bibinfo{volume}{18}},
  \bibinfo{pages}{023033} (\bibinfo{year}{2016}).

\bibitem[{\citenamefont{Schollw\"ock}(2011)}]{Sch11}
\bibinfo{author}{\bibfnamefont{U.}~\bibnamefont{Schollw\"ock}},
  \bibinfo{journal}{Ann. Phys. (N.Y.)} \textbf{\bibinfo{volume}{326}},
  \bibinfo{pages}{96} (\bibinfo{year}{2011}).

\bibitem[{\citenamefont{Zauner-Stauber
  et~al.}(2018)\citenamefont{Zauner-Stauber, Vanderstraeten, Fishman,
  Verstraete, and Haegeman}}]{ZSVF+18}
\bibinfo{author}{\bibfnamefont{V.}~\bibnamefont{Zauner-Stauber}},
  \bibinfo{author}{\bibfnamefont{L.}~\bibnamefont{Vanderstraeten}},
  \bibinfo{author}{\bibfnamefont{M.~T.} \bibnamefont{Fishman}},
  \bibinfo{author}{\bibfnamefont{F.}~\bibnamefont{Verstraete}},
  \bibnamefont{and} \bibinfo{author}{\bibfnamefont{J.}~\bibnamefont{Haegeman}},
  \bibinfo{journal}{Phys. Rev. B} \textbf{\bibinfo{volume}{97}},
  \bibinfo{pages}{045145} (\bibinfo{year}{2018}).

\bibitem[{\citenamefont{Hubig et~al.}(2015)\citenamefont{Hubig, McCulloch,
  Schollw\"ock, and Wolf}}]{HMSW15}
\bibinfo{author}{\bibfnamefont{C.}~\bibnamefont{Hubig}},
  \bibinfo{author}{\bibfnamefont{I.~P.} \bibnamefont{McCulloch}},
  \bibinfo{author}{\bibfnamefont{U.}~\bibnamefont{Schollw\"ock}},
  \bibnamefont{and} \bibinfo{author}{\bibfnamefont{F.~A.} \bibnamefont{Wolf}},
  \bibinfo{journal}{Phys. Rev. B} \textbf{\bibinfo{volume}{91}},
  \bibinfo{pages}{155115} (\bibinfo{year}{2015}).

\bibitem[{\citenamefont{McCulloch}(2007)}]{McC07}
\bibinfo{author}{\bibfnamefont{I.~P.} \bibnamefont{McCulloch}},
  \bibinfo{journal}{J. Stat. Mech. Theor. Exp.}
  \textbf{\bibinfo{volume}{2007}}, \bibinfo{pages}{P10014}
  (\bibinfo{year}{2007}).

\bibitem[{\citenamefont{Weichselbaum}(2012)}]{Wei12}
\bibinfo{author}{\bibfnamefont{A.}~\bibnamefont{Weichselbaum}},
  \bibinfo{journal}{Ann. Phys. (N.Y.)} \textbf{\bibinfo{volume}{327}},
  \bibinfo{pages}{2972} (\bibinfo{year}{2012}).

\bibitem[{\citenamefont{Luttinger and Ward}(1960)}]{LW60}
\bibinfo{author}{\bibfnamefont{J.~M.} \bibnamefont{Luttinger}}
  \bibnamefont{and} \bibinfo{author}{\bibfnamefont{J.~C.} \bibnamefont{Ward}},
  \bibinfo{journal}{Phys. Rev.} \textbf{\bibinfo{volume}{118}},
  \bibinfo{pages}{1417} (\bibinfo{year}{1960}).

\bibitem[{\citenamefont{Luttinger}(1960)}]{Lut60}
\bibinfo{author}{\bibfnamefont{J.~M.} \bibnamefont{Luttinger}},
  \bibinfo{journal}{Phys. Rev.} \textbf{\bibinfo{volume}{119}},
  \bibinfo{pages}{1153} (\bibinfo{year}{1960}).

\bibitem[{\citenamefont{Sayad et~al.}(2012)\citenamefont{Sayad, G\"utersloh,
  and Potthoff}}]{SGP12b}
\bibinfo{author}{\bibfnamefont{M.}~\bibnamefont{Sayad}},
  \bibinfo{author}{\bibfnamefont{D.}~\bibnamefont{G\"utersloh}},
  \bibnamefont{and} \bibinfo{author}{\bibfnamefont{M.}~\bibnamefont{Potthoff}},
  \bibinfo{journal}{Euro. Phys. J. B} \textbf{\bibinfo{volume}{85}},
  \bibinfo{pages}{125} (\bibinfo{year}{2012}).

\bibitem[{\citenamefont{Giuliani et~al.}(2005)\citenamefont{Giuliani, Vignale,
  and Datta}}]{GVD05}
\bibinfo{author}{\bibfnamefont{G.~F.} \bibnamefont{Giuliani}},
  \bibinfo{author}{\bibfnamefont{G.}~\bibnamefont{Vignale}}, \bibnamefont{and}
  \bibinfo{author}{\bibfnamefont{T.}~\bibnamefont{Datta}},
  \bibinfo{journal}{Phys. Rev. B} \textbf{\bibinfo{volume}{72}},
  \bibinfo{pages}{033411} (\bibinfo{year}{2005}).

\bibitem[{\citenamefont{Yafet}(1987)}]{Yaf87}
\bibinfo{author}{\bibfnamefont{Y.}~\bibnamefont{Yafet}},
  \bibinfo{journal}{Phys. Rev. B} \textbf{\bibinfo{volume}{36}},
  \bibinfo{pages}{3948} (\bibinfo{year}{1987}).

\bibitem[{chi()}]{chi0}
\bibinfo{note}{For the free electron gas in one dimension, one can evaluate
  Eq.\ (\ref{eq:chi0}) analytically and show that \cite{GVD05,Yaf87}
  $\chi_0(\omega=0, k) \sim \frac{\ln (2k_F-k)}{\ln (2k_F+k)}$. This diverges
  at $k=2k_F$. However, in the real-space representation one has
  $\chi_{ij}\sim\text{Si}(R_i-R_j)$, and the sinc function is integrable from 0
  to $\infty$ such that the total energy of a Hamiltonian of the form $H =
  L^{-1} \sum_{ij} \chi_{ij}S_iS_j$ stays finite for $L \to \infty$.}

\bibitem[{\citenamefont{Mermin and Wagner}(1966)}]{MW66}
\bibinfo{author}{\bibfnamefont{N.~D.} \bibnamefont{Mermin}} \bibnamefont{and}
  \bibinfo{author}{\bibfnamefont{H.}~\bibnamefont{Wagner}},
  \bibinfo{journal}{Phys. Rev. Lett.} \textbf{\bibinfo{volume}{17}},
  \bibinfo{pages}{1133} (\bibinfo{year}{1966}).

\bibitem[{\citenamefont{Liang}(1990)}]{Lia90}
\bibinfo{author}{\bibfnamefont{S.}~\bibnamefont{Liang}},
  \bibinfo{journal}{Phys. Rev. Lett.} \textbf{\bibinfo{volume}{64}},
  \bibinfo{pages}{1597} (\bibinfo{year}{1990}).

\bibitem[{\citenamefont{Lin and Campbell}(1991)}]{LC91}
\bibinfo{author}{\bibfnamefont{H.~Q.} \bibnamefont{Lin}} \bibnamefont{and}
  \bibinfo{author}{\bibfnamefont{D.~K.} \bibnamefont{Campbell}},
  \bibinfo{journal}{J. Appl. Phys.} \textbf{\bibinfo{volume}{69}},
  \bibinfo{pages}{5947} (\bibinfo{year}{1991}).

\bibitem[{\citenamefont{Sandvik and Scalapino}(1993)}]{SS93}
\bibinfo{author}{\bibfnamefont{A.~W.} \bibnamefont{Sandvik}} \bibnamefont{and}
  \bibinfo{author}{\bibfnamefont{D.~J.} \bibnamefont{Scalapino}},
  \bibinfo{journal}{Phys. Rev. B} \textbf{\bibinfo{volume}{47}},
  \bibinfo{pages}{12333} (\bibinfo{year}{1993}).

\bibitem[{\citenamefont{Koma and Mizukoshi}(1996)}]{KM96}
\bibinfo{author}{\bibfnamefont{T.}~\bibnamefont{Koma}} \bibnamefont{and}
  \bibinfo{author}{\bibfnamefont{N.}~\bibnamefont{Mizukoshi}},
  \bibinfo{journal}{J. Stat. Phys.} \textbf{\bibinfo{volume}{83}},
  \bibinfo{pages}{661} (\bibinfo{year}{1996}).

\bibitem[{\citenamefont{Yu and White}(1993)}]{YW93}
\bibinfo{author}{\bibfnamefont{C.~C.} \bibnamefont{Yu}} \bibnamefont{and}
  \bibinfo{author}{\bibfnamefont{S.~R.} \bibnamefont{White}},
  \bibinfo{journal}{Phys. Rev. Lett.} \textbf{\bibinfo{volume}{71}},
  \bibinfo{pages}{3866} (\bibinfo{year}{1993}).

\bibitem[{\citenamefont{Tsvelik}(1994)}]{Tsv94}
\bibinfo{author}{\bibfnamefont{A.~M.} \bibnamefont{Tsvelik}},
  \bibinfo{journal}{Phys. Rev. Lett.} \textbf{\bibinfo{volume}{72}},
  \bibinfo{pages}{1048} (\bibinfo{year}{1994}).

\bibitem[{\citenamefont{Wang et~al.}(1993)\citenamefont{Wang, Li, and
  Lee}}]{WLL93}
\bibinfo{author}{\bibfnamefont{Z.}~\bibnamefont{Wang}},
  \bibinfo{author}{\bibfnamefont{X.-P.} \bibnamefont{Li}}, \bibnamefont{and}
  \bibinfo{author}{\bibfnamefont{D.-H.} \bibnamefont{Lee}},
  \bibinfo{journal}{Phys. Rev. B} \textbf{\bibinfo{volume}{47}},
  \bibinfo{pages}{11935} (\bibinfo{year}{1993}).

\bibitem[{\citenamefont{Jullien and Pfeuty}(1981)}]{JP81}
\bibinfo{author}{\bibfnamefont{R.}~\bibnamefont{Jullien}} \bibnamefont{and}
  \bibinfo{author}{\bibfnamefont{P.}~\bibnamefont{Pfeuty}},
  \bibinfo{journal}{Journal of Physics F: Metal Physics}
  \textbf{\bibinfo{volume}{11}}, \bibinfo{pages}{353} (\bibinfo{year}{1981}).

\bibitem[{\citenamefont{Dublenych}(2016)}]{Dub16}
\bibinfo{author}{\bibfnamefont{Y.~I.} \bibnamefont{Dublenych}},
  \bibinfo{journal}{Phys. Rev. B} \textbf{\bibinfo{volume}{93}},
  \bibinfo{pages}{054415} (\bibinfo{year}{2016}).

\bibitem[{\citenamefont{Lieb et~al.}(1961)\citenamefont{Lieb, Schultz, and
  Mattis}}]{LSM61}
\bibinfo{author}{\bibfnamefont{E.}~\bibnamefont{Lieb}},
  \bibinfo{author}{\bibfnamefont{T.}~\bibnamefont{Schultz}}, \bibnamefont{and}
  \bibinfo{author}{\bibfnamefont{D.}~\bibnamefont{Mattis}},
  \bibinfo{journal}{Ann. Phys. (N.Y.)} \textbf{\bibinfo{volume}{16}},
  \bibinfo{pages}{407} (\bibinfo{year}{1961}).

\bibitem[{\citenamefont{Majumdar and Ghosh}(1969)}]{MG69}
\bibinfo{author}{\bibfnamefont{C.~K.} \bibnamefont{Majumdar}} \bibnamefont{and}
  \bibinfo{author}{\bibfnamefont{D.~K.} \bibnamefont{Ghosh}},
  \bibinfo{journal}{J. Math. Phys.} \textbf{\bibinfo{volume}{10}},
  \bibinfo{pages}{1388} (\bibinfo{year}{1969}).

\bibitem[{\citenamefont{Haldane}(1983{\natexlab{a}})}]{Hal83a}
\bibinfo{author}{\bibfnamefont{F.~D.~M.} \bibnamefont{Haldane}},
  \bibinfo{journal}{Phys. Lett. A} \textbf{\bibinfo{volume}{93}},
  \bibinfo{pages}{464} (\bibinfo{year}{1983}{\natexlab{a}}).

\bibitem[{\citenamefont{Haldane}(1983{\natexlab{b}})}]{Hal83b}
\bibinfo{author}{\bibfnamefont{F.~D.~M.} \bibnamefont{Haldane}},
  \bibinfo{journal}{Phys. Rev. Lett.} \textbf{\bibinfo{volume}{50}},
  \bibinfo{pages}{1153} (\bibinfo{year}{1983}{\natexlab{b}}).

\end{thebibliography}
\end{document}